\documentclass[journal=jctcce,manuscript=article,layout=twocolumn]{achemso}
\usepackage{amsmath}
\usepackage{amsfonts}
\usepackage{graphicx}
\usepackage{color} 
\usepackage{subcaption}
\usepackage{braket}
\usepackage{cleveref}
\usepackage{natbib}

\usepackage{math_abbreviations}

\graphicspath{{figures/}}

\author{Stefano Battaglia}
\email{stefano.battaglia@kemi.uu.se}

\author{Roland Lindh}
\email{roland.lindh@kemi.uu.se}

\affiliation{
  Department of Chemistry - BMC,
  Uppsala University,
  P.O. Box 576,
  SE-75123 Uppsala,
  Sweden
}

\title{Extended dynamically weighted CASPT2:\newline the best of two worlds}


\begin{document}

\begin{abstract}
  We introduce a new variant of the complete active space second-order perturbation theory (CASPT2)
  method that performs similarly to multistate CASPT2 (MS-CASPT2) in regions of the potential energy
  surface where the electronic states are energetically well separated and is akin to extended
  MS-CASPT2 (XMS-CASPT2) in case the underlying zeroth-order references are near-degenerate.
  Our approach follows a recipe analogous to XMS-CASPT2 to ensure approximate invariance under
  unitary transformations of the model states and a dynamical weighting scheme to smoothly
  interpolate the Fock operator between a state-specific and a state-average regime.
  The resulting extended dynamically weighted CASPT2 (XDW-CASPT2) methodology possesses the most
  desirable features of both MS-CASPT2 and XMS-CASPT2, i.e. the ability to provide accurate
  transition energies and correctly describe avoided crossings and conical intersections.
  The reliability of XDW-CASPT2 is assessed on a number of molecular systems.
  First, we consider the dissociation of lithium fluoride, highlighting the distinctive
  characteristics of the new approach. Second, the invariance of the theory is investigated
  by studying the conical intersection of the distorted allene molecule.
  Finally, the relative accuracy in the calculation of vertical excitation energies is benchmarked
  on a set of 26 organic compounds.
  We found that XDW-CASPT2, albeit being only approximately invariant, produces smooth potential
  energy surfaces around conical intersections and avoided crossings, performing equally well to
  the strictly invariant XMS-CASPT2 method.
  The accuracy of vertical transition energies is almost identical to MS-CASPT2, with a mean
  absolute deviation of 0.01 to 0.02 eV, in contrast to 0.12 eV for XMS-CASPT2.
\end{abstract}

\section{Introduction}

The theoretical modeling of excited states processes is undoubtedly of fundamental and
practical importance\cite{Matsika2018}.
The investigation of physical mechanisms at the base of chemi- and bioluminescence\cite{Vacher2018},
spectroscopy\cite{Norman2018}, singlet fission\cite{Casanova2018}, and many other scientifically and
technologically relevant applications require methodologies that are able to describe the entire
potential energy surface (PES), providing accurate relative energies between different electronic
states, their correct orderings and the right morphology in regions of
near-degeneracies\cite{Lischka2018}.
Single-reference approaches, despite their widespread success, do not generally have
the necessary flexibility to be applied indistinctly in any situation that one might
encounter in the realm of excited states chemistry: a multireference approach is unavoidable.
From the several available options, multireference perturbation
theory (MRPT) stands out: its accuracy, general applicability and moderate computational cost
elected it during the last few decades as the method of choice for the investigation of full
potential energy surfaces\cite{Lischka2018}.
In particular, formalisms that allow the relaxation of the reference states under the influence
of the perturbation have seen the most success\cite{Cave1988,Nakano1993,Malrieu1995,Hoffmann1993,
Hoffmann1996,Finley1998,Mahapatra1999,Shavitt2002,Angeli2004,Fink2009,Roskop2011,Sharma2016,Giner2017,
Garniron2018,Bozkaya2019}, with the multistate complete active space second-order perturbation
theory (MS-CASPT2)\cite{Finley1998} approach being one of the most popular.
Relying on the multipartitioning technique\cite{Malrieu1995}, this methodology is well suited for
the calculation of transition energies between states that are well separated, with deviations
within 0.1 to 0.2 eV from the best theoretical estimates\cite{Schreiber2008,Silva-Junior2010}.
On the other hand, even though MS-CASPT2 follows the ``diagonalize-then-perturb-then-diagonalize''
philosophy, it may still suffer from unphysical behaviors at molecular geometries with near-degenerate
reference states.
The theoretical understanding of this shortcoming is known\cite{Granovsky2011} and it can be solved
by enforcing the states to be invariant under unitary transformations within the
model space, leading to the so-called extended MS-CASPT2 (XMS-CASPT2)\cite{Shiozaki2011} method.
However, the latter requires a unique partitioning of the Hamiltonian achieved through the use of a
state-average Fock operator, which is likely to degrade the accuracy as the dimension of the model
space is increased or when the states under consideration are of different character
(e.g. valence and Rydberg).

The main objective of this work is to formulate a new CASPT2 variant that retains the accuracy of
MS-CASPT2 in the calculation of transition energies and at the
same time yields smooth potential energy surfaces with no artifacts in regions where the excited
states manifold is near-degenerate.
Our approach is based on the same transformation carried out in the initial step of XMS-CASPT2
and then uses a dynamical weighting scheme to interpolate between state-specific
and state-average operators; hence we call it extended dynamically weighted CASPT2 (XDW-CASPT2).
From a theoretical standpoint, XDW-CASPT2 corresponds to a new and somewhat sophisticated
partitioning of the Hamiltonian, thus retaining the underlying structure of the parent theory.
Recently, an analogous attempt to find a zeroth-order Hamiltonian that coincides with the canonical
MS-CASPT2 one, but that at the same time is invariant as in XMS-CASPT2 was carried out by
\citet{Park2019}.
XDW-CASPT2 also shares some similarities with the recently introduced dynamically weighted
driven similarity renormalization group (DW-DSRG)\cite{Li2019a} as well as the dynamically weighted
complete active space self-consisted field method\cite{Deskevich2004,Glover2014}.

The article is structured as follows.
In section 2 we first selectively review important aspects of quasidegenerate perturbation theory
(QDPT), MS-CASPT2 and XMS-CASPT2 necessary to define XDW-CASPT2 in the remainder of the section.
Next, section 3 is devoted to the assessment of the new methodology and is divided in three parts.
First, an extensive study on the dissociation of LiF is presented, being this problem a prototypical
example to show all features of the new method. Second, the conical intersection in the distorted
allene molecule is investigated, which represents a difficult case for QDPT-based approaches.
Third, the accuracy of vertical transition energies to the lowest singlet excited state
is evaluated on a set of 26 small to medium organic compounds.
At last, in section 4 we conclude by summarizing the results obtained in this contribution and
with an outlook on future directions regarding XDW-CASPT2.

\section{Theory}

As in any perturbation theory approach, the starting point is to partition the full Hamiltonian
into a zeroth-order part $\hH_0$, with known eigenfunctions $\Psi^{(0)}_{\alpha}$ and associated
eigenvalues $E^{(0)}_{\alpha}$, and a perturbation operator $\hV = \hH - \hH_0$. 
The Hilbert space is also partitioned into a model space, spanned by model functions (also called
reference functions) selected from the set of zeroth-order ones, and a complementary space, spanned
by all other functions orthogonal to the model ones.
The projector onto the model space is defined as
\begin{equation}
  \label{eq:P_projector}
  \hP = \sum_{\gamma\in\mcP}\kPzg\bPzg
\end{equation}
where $\mcP$ denotes the set of indices labeling the reference states.
The projector onto the complementary space is simply defined as $\hQ = \hat{1} - \hP$.
Note that the complementary space does not necessarily have to be spanned by the remaining
zeroth-order functions not included in the model space: other types of many-electron functions
can be used.
The wave operator\cite{Bloch1958} $\hat{\Omega}$, defined as an operator that acting on a
model state $\Pza$ generates the exact one (i.e. of the full Hamiltonian $\hH$)
\begin{equation}
  \label{eq:wave_operator}
  \hat{\Omega}\Pza = \Psi_{\alpha}
\end{equation}
is governed by the generalized Bloch equation\cite{Lindgren1974}
\begin{equation}
  \label{eq:Bloch}
  \lbrack\hat{\Omega},\hH_0\rbrack = \hQ\hV\hat{\Omega} - \hQ\hat{\Omega}\hV\hat{\Omega}
\end{equation}
Assuming intermediate normalization, an effective Hamiltonian is constructed according to
\begin{equation}
  \label{eq:Heff}
  \hH_{eff} = \hP\hH\hat{\Omega}\hP
\end{equation}
whose eigenvalues and eigenfunctions (within the model space) correspond to the exact ones.
To arrive at a practical implementation of QDPT, $\hat{\Omega}$ is expanded in powers of the
perturbation operator
\begin{align}
  \label{eq:Omega_expansion}
  \hat{\Omega} = \hat{\Omega}^{(0)} + \hat{\Omega}^{(1)} + \hat{\Omega}^{(2)} + \ldots
\end{align}
and substituted into \Cref{eq:Heff} leading to the second-order effective Hamiltonian
\begin{equation}
  \label{eq:Heff_2nd}
  \hH_{eff}^{(2)} = \hP\hH\hat{\Omega}^{(1)}\hP
\end{equation}
with the superscript ${(n)}$ denoting the order in $\hV$.
Expressing $\hH_{eff}^{(2)}$ in the model space basis and diagonalizing the resulting matrix
provides the second-order correction to the energies and the perturbatively modified
zeroth-order wave functions.
To determine $\hat{\Omega}^{(1)}$ in \Cref{eq:Heff_2nd}, one has to solve the first-order generalized
Bloch equation
\begin{equation}
  \label{eq:Bloch_1st}
  \lbrack\hat{\Omega}^{(1)},\hH_0\rbrack = \hQ\hV\hP
\end{equation}
which is obtained upon inserting \Cref{eq:Omega_expansion}
into \Cref{eq:Bloch} and equating only the terms which are of first order in $\hV$.
Note that the application of $\hat{\Omega}^{(1)}$ to $\Pza$ generates the first-order correction
to the wave function
\begin{equation}
  \label{eq:Psi_1}
  \Poa = \hat{\Omega}^{(1)}\Pza
\end{equation}
for all $\alpha \in \mcP$.

\subsection{MS-CASPT2}

In the multistate CASPT2 method\cite{Finley1998} the zeroth-order functions defining the model space
are of complete active space self-consistent field (CASSCF) type. For each $\alpha \in \mcP$, there
is a separate partitioning of the full Hamiltonian\cite{Zaitsevskii1995},
$\hH = \hH_0^{\alpha} + \hV^{\alpha}$, with the zeroth-order part defined by
\begin{equation}
  \label{eq:H0_MS}
  \begin{aligned}
    \hH_0^{\alpha} &=  \sum_{\gamma\in\mcP} \kPzg\bPzg\hf^{\alpha}\kPzg\bPzg \\
    &+ \sum_{k\in\mcP^{\perp}} \ket{\Pz_k}\bra{\Pz_k}\hf^{\alpha}\ket{\Pz_k}\bra{\Pz_k} \\
    &+ \hQ_{SD} \hf^{\alpha} \hQ_{SD} + \hQ_{TQ\ldots} \hf^{\alpha} \hQ_{TQ\ldots}
  \end{aligned}
\end{equation}
The first sum is restricted to states in the model space (including $\gamma = \alpha$), while the
second one runs over all other states of the complete active space, with $\mcP^{\perp}$ being
the set of indices labeling them.
The remaining part of the complementary space is spanned by internally contracted configurations
(ICCs) obtained by the application of excitation operators to the reference states $\Pza$.
The operator $\hQ_{SD}$ projects onto the so-called first-order interacting space that, for the sake
of this theoretical discussion, we shall assume is always generated from the union of all model
states\footnote{This seemingly
insignificant choice avoids a discussion on single-state single-reference (SS-SR) and multistate
multireference (MS-MR) variants of internally contracted theories which is not the primary focus
of this work. For a thorough comparison between them in the context of CASPT2 see e.g. the recent
work by \citet{Park2019}.}.
Similarly, $\hQ_{TQ\ldots}$ projects onto the space spanned by higher-order ICCs.
The generalized Fock operator $\hf^{\alpha}$ is given by
\begin{equation}
  \label{eq:f_MS}
  \hf^{\alpha} = \sum_{pq} f_{pq}^{\alpha} \hE_{pq}
\end{equation}
where $\hE_{pq}$ is the second-quantized spin-summed one-particle excitation operator
and $f_{pq}^{\alpha}$ are entries of the Fock matrix expressed in the molecular orbital basis
\begin{equation}
  \label{eq:fpq_MS}
  f_{pq}^{\alpha} = h_{pq} + \sum_{rs} D_{rs}^{\alpha} \left[ (pq|rs) - \frac{1}{2}(pr|qs) \right]
\end{equation}
Here, $h_{pq}$ and $(pq|rs)$ are elements of the one-particle Hamiltonian and the two-electron
repulsion integrals, respectively, while $D_{rs}^{\alpha}=\braket{\Pza|\hE_{rs}|\Pza}$ are entries of
the one-particle reduced density matrix (1-RDM, or also simply called density matrix) of state $\Pza$.
The indices $p,q,r$ and $s$ label general molecular orbitals.
The use of projectors in \Cref{eq:H0_MS} for the definition of $\hH_0$ is necessary because the
CASSCF states are not eigenfunctions of the generalized Fock operator.
Furthermore, note that $\hH_0$ is diagonal within the model space since $\hf^{\alpha}$ is projected
directly onto the reference states rather than onto the space spanned by them. In other words,
even though in general
\begin{equation}
  \label{eq:fab_MS_neq_zero}
  \braket{\Pza|\hf^{\gamma}|\Pzb} \neq 0
\end{equation}
for $\alpha \neq \beta \in \mcP$ and $\gamma \in \mcP$, these elements are arbitrarily set to zero
in the MS-CASPT2 $\hH_0$.
This constitutes an approximation that we will call hereafter \emph{diagonal approximation}.
The immediate consequence of this choice is that upon inserting \Cref{eq:H0_MS} into
\Cref{eq:Bloch_1st}, the solution of the first-order generalized Bloch equation can be obtained for
each state of the model space separately, as these are not coupled anymore.
The substantial advantage gained is the possibility to use state-specific Fock operators
in $\hH_0$, allowing for a formalism based on multipartitioning that
should provide more accurate zeroth-order energies. In particular, for states that are
energetically well separated or have considerably different character, state-specific Fock operators
are in principle better suited to describe them than, for instance, a single operator that
requires the flexibility to account for all states in an average way.

On the other hand, the diagonal approximation has a profound impact on the invariance properties
of the method as elucidated by \citet{Granovsky2011} in the context of multiconfigurational QDPT
(MCQDPT).
The main issues are \emph{two}. First, when two reference states interact strongly at zeroth-order,
meaning that the element $\braket{\Pza|\hf^{\gamma}|\Pzb}$ is significantly larger than zero, it can
be shown\cite{Granovsky2011} that neglecting it leads to large systematic errors in the corresponding
off-diagonal element of the second-order effective Hamiltonian.
Second, it is known that zeroth-order states at a conical intersection (CI), and to a large extent
at an avoided crossing (AC) as well, are not well-defined: in such situations any linear superposition
of the involved states constitutes an equally valid wave function.
Hence, simply projecting the Fock operator onto the individual components entails an arbitrary
choice which might lead to the appearance of artifacts on the potential energy surface in the
vicinity of the AC or CI.

In conclusion, we should note that it is possible to adopt a unique partitioning in MS-CASPT2,
for instance with the use of a state-average Fock operator. Such an approach however would lose
the advantages of multipartitioning, but keep the issues related to the lack of invariance.
Nevertheless, this strategy has been recently explored by \citet{Kats2019} in the context of
pair natural orbital MS-CASPT2, finding systematic deviations from canonical MS-CASPT2 by 0.1 to
0.2 eV for transitions to the lowest singlet excited state.

\subsection{XMS-CASPT2}

The main flaw of MS-CASPT2 is the lack of invariance under unitary transformations within
the model space. The result obtained with a particular set of reference states should always be
the same to the one obtained with a set of states generated by a unitary transformation of the
original ones.
This shortcoming is ascribed to the diagonal approximation of $\hH_0$ and the solution to this
problem was first proposed by \citet{Granovsky2011} for MCQDPT and shortly after applied to
MS-CASPT2 by \citet{Shiozaki2011}.
The key difference of the new methodology, XMS-CASPT2, is in the zeroth-order Hamiltonian
\begin{equation}
  \label{eq:H0_XMS}
  \begin{aligned}
    &\hH_0 =  \sum_{\gamma,\delta\in\mcP} \kPzg\bPzg\hf^{sa}\kPzd\bPzd \\
    &+ \sum_{k\in\mcP^{\perp}} \ket{\Pz_k}\bra{\Pz_k}\hf^{sa}\ket{\Pz_k}\bra{\Pz_k} \\
    &+ \hQ_{SD} \hf^{sa} \hQ_{SD} + \hQ_{TQ\ldots} \hf^{sa} \hQ_{TQ\ldots}
  \end{aligned}
\end{equation}
The Fock operator is now projected onto the full model space rather than on the individual
components alone. This implies a unique partitioning of the Hamiltonian because the first-order
generalized Bloch equation, \Cref{eq:Bloch_1st}, does not decouple the states anymore.
The Fock operator $\hf^{sa}$ is constructed from the state-average density matrix
\begin{equation}
  \label{eq:D_sa}
  \mbf{D}^{sa} = \frac{1}{d} \sum_{\alpha\in\mcP} \mbf{D}^{\alpha}
\end{equation}
for a model space containing $d$ states\footnote{Note that using the state-average density matrix
to construct the Fock operator or averaging the state-specific Fock operators lead to the same
$\hf^{sa}$.}.
The fact that $\hH_0$ is no longer diagonal in the zeroth-order basis makes the solution of
\Cref{eq:Bloch_1st} somewhat harder.
However, this complication can be fully overcome by a unitary transformation of the reference states,
such that the rotated wave functions
\begin{equation}
  \label{eq:Psi_0_rotated}
  \Pzta = \sum_{\beta\in\mcP} U_{\beta\alpha} \Pzb
\end{equation}
diagonalize the Fock operator within the model space. In other words, the rotated model states satisfy
\begin{equation}
  \label{eq:fab_XMS_eq_zero}
  \braket{\Pzta|\hf^{sa}|\Pztb} = 0
\end{equation}
for $\alpha\neq\beta \in \mcP$.
Using the wave functions $\Pzta$, the zeroth-order Hamiltonian can now be rewritten as
\begin{equation}
  \label{eq:H0_XMS_rotated}
  \begin{aligned}
    &\hH_0 =  \sum_{\gamma\in\mcP} \kPztg\bPztg\hf^{sa}\kPztg\bPztg \\
    &+ \sum_{k\in\mcP^{\perp}} \ket{\Pz_k}\bra{\Pz_k}\hf^{sa}\ket{\Pz_k}\bra{\Pz_k} \\
    &+ \hQ_{SD} \hf^{sa} \hQ_{SD} + \hQ_{TQ\ldots} \hf^{sa} \hQ_{TQ\ldots}
  \end{aligned}
\end{equation}
which has the same form of \Cref{eq:H0_MS}, albeit the use of the state-average Fock operator.
Therefore, MS-CASPT2 truly corresponds to an approximation of XMS-CASPT2, provided that
the same unique partitioning of the Hamiltonian is used in both variants.
The generalizations introduced with \Cref{eq:H0_XMS} make this method invariant under unitary
transformations of the model space wave functions, solving the issues intrinsic to the diagonal
approximation of MS-CASPT2.
As a result, XMS-CASPT2 is more robust in general, with energies that are continuous
and smooth functions of the molecular geometry even in the vicinity of ACs and CIs.
The price to pay is the use of $\hf^{sa}$ in $\hH_0$, which, as briefly discussed at the end of the
previous subsection, is likely to decrease the accuracy of the method for the calculation of
excitation energies, in particular when these are large or between states of different character.
Nevertheless, to the best of our knowledge, no systematic and comprehensive benchmark on the
accuracy of XMS-CASPT2 is available in the literature.

\subsection{XDW-CASPT2}

The necessary ingredients to design a hybrid approach that interpolates between MS-CASPT2 and
XMS-CASPT2 are the use of state-specific Fock operators in a multipartitioning formalism and the
projection of $\hH_0$ onto the full model space rather than onto individual reference states.
The objective is a method
that performs as well as MS-CASPT2 in situations where states are clearly discernible
and is as robust as XMS-CASPT2 when these are instead quasidegenerate.
We note from our previous discussion that in case the zeroth-order Hamiltonian has negligible
off-diagonal elements within the model space, i.e.
\begin{equation}
  \label{eq:H0ab_approx_0}
  \braket{\Pza|\hH_0|\Pzb} \approx 0
\end{equation}
the diagonal approximation is a sound simplification of the generalized Bloch equation.
Crucially, it allows for a formalism based on multipartitioning.
Thus, we are seeking a unitary transformation as in \Cref{eq:Psi_0_rotated}, whereby
the rotated states satisfy
\begin{equation}
  \label{eq:fab_XDW_approx_0}
  \braket{\Pzta|\hfb^{\gamma}|\Pztb} \approx 0
\end{equation}
for $\alpha\neq\beta \in \mcP$, with the Fock operator $\hfb^{\gamma}$ (note the bar to differentiate
this operator from the normal state-specific one) having the following property
\begin{equation}
  \label{eq:f_XDW_properties}
  \hfb^{\gamma} \approx
  \begin{cases}
    \hf^{\gamma} & \text{if $\Pztg$ weakly interacts} \\ & \text{with other model states} \\
    \hf^{sa}     & \text{if $\Pztg$ strongly interacts} \\ & \text{with other model states}
  \end{cases}
\end{equation}
for all $\gamma \in \mcP$.
Note that we shall better specify further below what does weak and strong interaction mean in this
context.

We are able to satisfy \Cref{eq:fab_XDW_approx_0,eq:f_XDW_properties} with the following scheme.
In a first step, completely analogous to XMS-CASPT2, a set of rotated model states $\Pzta$
is obtained by diagonalization of the state-average Fock operator $\hf^{sa}$.
These functions are then used to construct dynamically weighted density matrices of the form
\begin{equation}
  \label{eq:D_XDW}
  \mbf{\bar{D}}^{\alpha} = \sum_{\beta\in\mcP} \omega_{\alpha}^{\beta} \mbf{\tilde{D}}^{\beta}
\end{equation}
with weights satisfying the condition
\begin{equation}
  \label{eq:weight_condition}
  \sum_{\beta\in\mcP} \omega_{\alpha}^{\beta} = 1
\end{equation}
for all $\alpha\in\mcP$. The use of tildes emphasizes that $\mbf{\tilde{D}}^{\beta}$ is the
1-RDM associated to the rotated state $\Pztb$.
Using the densities defined in \Cref{eq:D_XDW}, state-specific Fock operators are constructed
according to \Cref{eq:f_MS,eq:fpq_MS} for all $\alpha \in \mcP$ and used to define the partitionings
of the Hamiltonian for a subsequent MS-CASPT2 calculation.
Thus, XDW-CASPT2 substantially consists in a MS-CASPT2 calculation employing zeroth-order states
defined by \Cref{eq:Psi_0_rotated} and state-specific Fock operators constructed with densities
$\mbf{\bar{D}}^{\alpha}$.

The weights $\omega_{\alpha}^{\beta}$ are chosen such that the resulting Fock operators
$\hfb^{\alpha}$ satisfy the prescription of \Cref{eq:f_XDW_properties}.
This is achieved by using a scheme recently introduced by one of the authors of this contribution
and his collaborators\cite{Li2019a}, whereby $\omega_{\alpha}^{\beta}$ is defined by the following
Boltzmann-like function
\begin{equation}
  \label{eq:boltzmann_weight}
  \omega_{\alpha}^{\beta} =
  \frac{e^{-\zeta \left( \Delta_{\alpha\beta} \right)^2}}
  {\sum_{\gamma\in\mcP} e^{-\zeta \left( \Delta_{\alpha\gamma} \right)^2}}
\end{equation}
where $\Delta_{\alpha\beta}$ ($\Delta_{\alpha\gamma}$) quantifies the interaction between
states $\Pzta$ and $\Pztb$ ($\Pztg$) and $\zeta \in \mathbb{R}^+_0$ is a parameter controlling the
sharpness of the transition between a mixed-density and a state-specific regime.
Let us list the asymptotic properties of \Cref{eq:boltzmann_weight} with respect to
$\Delta_{\alpha\beta}$:
\begin{alignat}{2}
  \label{eq:Delta_to_infty}
  \Delta_{\alpha\beta} &\to \infty &&\implies \omega_{\alpha}^{\beta} \to 0 \\
  \label{eq:Delta_to_zero}
  \Delta_{\alpha\beta} &\to 0      &&\implies \omega_{\alpha}^{\beta}=\omega_{\beta}^{\alpha}
\end{alignat}
A physical quantity that satisfies \Cref{eq:Delta_to_infty,eq:Delta_to_zero} is given by the
energy difference between the rotated states
\begin{equation}
  \label{eq:Delta_ab}
  \Delta_{\alpha\beta} = |\braket{\Pzta|\hH|\Pzta} - \braket{\Pztb|\hH|\Pztb}|
\end{equation}
When computing the contribution of state $\Pzb$ to the density of $\Pza$, if their energy
difference is large --- $\Delta_{\alpha\beta} \gg 0$ --- then $\Pzb$ should not contribute:
$\omega_{\alpha}^{\beta} \approx 0$. This situation corresponds to the case in which the two
states are weakly or not interacting.
Vice-versa, if the energy difference is small --- $\Delta_{\alpha\beta} \approx 0$ --- then
$\Pzb$ is quasidegenerate with $\Pza$ and should receive approximately the same weight:
$\omega_{\alpha}^{\beta} \approx \omega_{\beta}^{\alpha}$. This situation corresponds to the
case in which the two states are strongly interacting.
Note that, somewhat counterintuitively, strong interaction is associated with a small value of
the parameter $\Delta_{\alpha\beta}$ and conversely weak interaction with a large one.
Simply using an energetic criterion to parametrize the interaction strength between
two states can lead to unphysical averaging: e.g. two states of different symmetry (spin or spatial)
should not be mixed together irrespective of their relative energy.
In this work, this problem has been circumvented by treating states of different symmetry separately,
however, a more general solution is possible.
For instance, by multiplying the right-hand side of \Cref{eq:Delta_ab} by a factor dependent on the
off-diagonal element of the full Hamiltonian expressed in the basis of rotated references,
$\braket{\Pzta|\hH|\Pztb}$, $\Delta_{\alpha\beta}$ would account for the physical nature of the
states without resorting on external constraints (e.g. forcing symmetries).
Importantly, such a modification would correctly model changes of the molecular geometry that break
the symmetry of the system.

For a fixed state $\Pza$, the parameter $\zeta$ modulates the importance of other states in a
collective manner: a small value tends to make them all equally important, whereas a large value
favors the parent state under consideration.
The asymptotic behavior of $\omega_{\alpha}^{\beta}$ with respect to $\zeta$ is given by
\begin{alignat}{2}
  \label{eq:zeta_to_infty}
  \zeta &\to \infty &&\implies
  \begin{cases}
  \omega_{\alpha}^{\beta} \to 0 & \text{if } \beta \neq \alpha \\
  \omega_{\alpha}^{\beta} \to 1 & \text{if } \beta = \alpha
  \end{cases} \\
  \label{eq:zeta_to_zero}
  \zeta &\to 0 &&\implies\omega_{\alpha}^{\beta}\to\frac{1}{d} \quad \forall \beta \in \mcP
\end{alignat}
where $d$ is the number of model states.
The situation depicted in \Cref{eq:zeta_to_infty} results in purely state-specific Fock operators
alike MS-CASPT2, albeit using densities of rotated reference functions.
On the other hand, if all weights are equal as shown in \Cref{eq:zeta_to_zero}, the
original XMS-CASPT2 is restored.
For the particular choice of $\Delta_{\alpha\beta}$ made in \Cref{eq:Delta_ab}, $\zeta$ assumes
$\Eh^{-2}$ units and its value can be regarded as a threshold. When the value of
$\Delta_{\alpha\beta}$ is in the same order of magnitude as $\zeta^{-1/2}$ or smaller, the state
$\Pzb$ will contribute significantly to $\hfb^{\alpha}$, if instead
$\Delta_{\alpha\beta} \gg \zeta^{-1/2}$, it will play little to no role in $\hfb^{\alpha}$.

As a final remark, we should note that \Cref{eq:boltzmann_weight} was used in a similar fashion in
the recent work by \citet{Li2019a},
where not only the 1-RDM was averaged with dynamical weights, but also higher-order RDMs.
Importantly, the latter were introduced in the flow equations, whose solution provides the diagonal
matrix elements of the effective Hamiltonian. In the approach presented here, the densities defined
by \Cref{eq:D_XDW} are only used to obtain an alternative partitioning of the Hamiltonian;
the first-order equations that determine the correction to the wave function, and accordingly
$\hH_{eff}$, make use of purely state-specific densities.

\section{Results}

In this section we present the results obtained for a series of calculations representing typical
use-case scenarios in order to assess the reliability of XDW-CASPT2.
First, the avoided crossings in LiF are investigated. This prototypical system is an ideal model
to highlight the strengths, weaknesses and features of XDW-CASPT2 as compared to MS-CASPT2 and
XMS-CASPT2. Since this example touches every aspect of the theory, the discussion of this case is
quite extensive.
Second, the conical intersection in the distorted allene molecule is considered. This system
provides a tougher test for the invariance properties of the theory, thereby probing the robustness
of the approach.
At last, singlet vertical excitation energies are computed for a series of organic compounds in
order to evaluate the accuracy of the method and the effect of the dynamical weighting scheme.

All calculations were performed with a development branch of OpenMolcas\cite{OpenMolcas} based on
the master branch, version v18.09-617-g5a96a25e.
Note that the CASPT2 implementation of OpenMolcas uses the SS-SR ICC basis, thereby never fully
preserving invariance, not even for XMS-CASPT2.

\subsection{Avoided crossings in LiF}

It is well known that during the dissociation of lithium fluoride the two lowest singlet states
of $^1\Sigma^+$ symmetry undergo a rapid change of character switching between ionic and covalent.
A state-average CASSCF (SA-CASSCF) calculation predicts the avoided crossing at a much shorter
distance compared to the reference values (e.g. full configuration interaction) due to the missing
dynamical electron correlation\cite{Bauschlicher1988,Finley1998}.
Introduction of the latter in a state-specific manner, for instance through single-state CASPT2,
results in an artificial double crossing of the two potential energy curves (PECs), which, alongside
other issues present in the theory, has been a main motivation for the development of its
multistate generalization.
Nevertheless, even though MS-CASPT2 provides much more satisfactory results, it still faces severe
complications at internuclear distances where the underlying reference
states are quasidegenerate. This is particularly visible when considering the three lowest
$^1\Sigma^+$ states rather than the usual two.
Instead, XMS-CASPT2 does not incur in any unphysical behavior irrespective of the number of states,
however at the expense of a reduced accuracy in their relative energy at the equilibrium distance.
Thus, lithium fluoride is an ideal system to test XDW-CASPT2, and to this end, we calculated
its dissociation considering the three lowest singlet states simultaneously.

\begin{figure}
  \centering
  \includegraphics{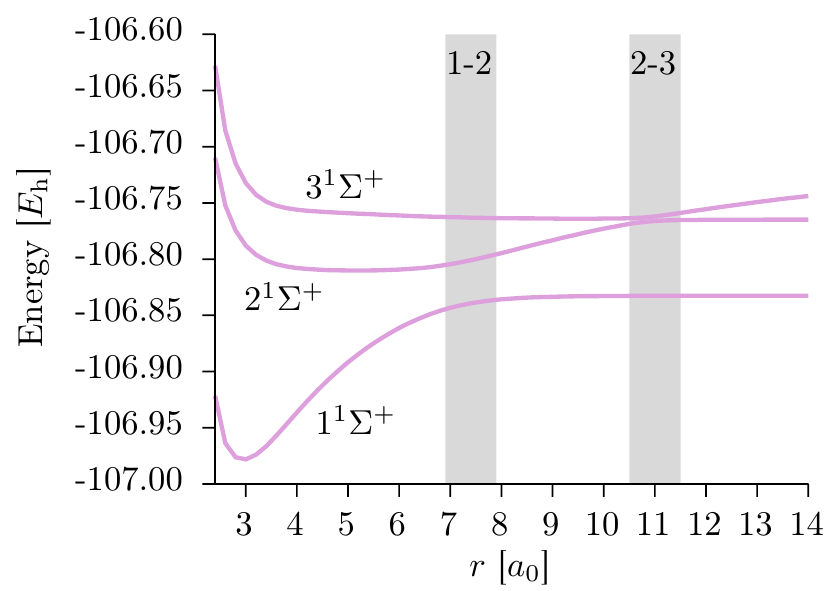}
  \caption{SA-CASSCF potential energy curves of the three lowest $^1\Sigma^+$ states of lithium
  fluoride. There are two avoided crossing regions (highlighted in gray), one between the
  ground and the first excited state, labeled 1-2, and one between the first and the second
  excited state, labeled 2-3.}
  \label{fig:LiF_SA3-CASSCF}
\end{figure}

The reference wave functions were obtained by a SA-CASSCF\cite{Roos1980} calculation using equal
weights for all three states and imposing the C$_{2v}$ molecular point group symmetry.
The active space was composed by six electrons in 2 a$_1$, 2 b$_1$ and 2 b$_2$ orbitals, while the
remaining 3 occupied a$_1$ orbitals were relaxed during optimization.
The cc-pVTZ\cite{DunningJr1989} and aug-cc-pVTZ basis set\cite{Kendall1992} were
used on lithium and fluorine, respectively.
The potential energy curves were computed for internuclear distances between 2.4 and 14 $a_0$
in steps of 0.2 $a_0$.
The results obtained with the CASSCF method are shown in \Cref{fig:LiF_SA3-CASSCF}.
At an internuclear distance comprised between 6.8 and 7.8 $a_0$, the ground state wave function
quickly changes from an ionic to a covalent character, whereas the opposite happens for the
$2^1\Sigma^+$ one. The inclusion of a third state in the calculation plays a little
role here: the position of this avoided crossing is slightly shifted to a shorter internuclear
distance compared to a 2-state calculation (see Supporting Information for 2-state PECs).
From 10.5 to 11.5 $a_0$, a second avoided crossing between the $2^1\Sigma^+$ and the
$3^1\Sigma^+$ states appear, where the character of the $3^1\Sigma^+$ wave function becomes
ionic.
These quasidegeneracies among the CASSCF states have important consequences on the accuracy
and effectiveness of the perturbative approach used to recover the dynamic electron
correlation.

\begin{figure}
  \centering
  \includegraphics{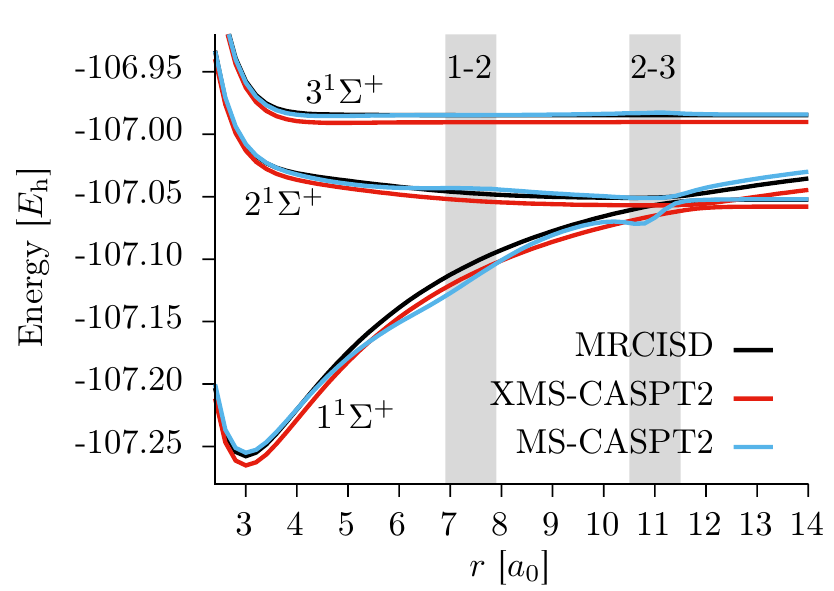}
  \caption{Potential energy curves of the three lowest $^1\Sigma^+$ states of lithium
  fluoride. The zones highlighted in gray correspond to the avoided crossing regions
  at the CASSCF level of theory.}
  \label{fig:LiF_reference}
\end{figure}

To establish a reference, we report in \Cref{fig:LiF_reference} the PECs computed with MS-CASPT2,
XMS-CASPT2 and multireference configuration interaction with singles and doubles (MRCISD).
For all three methodologies, the two 1s core orbitals were kept frozen, and the 2s orbital
of fluorine was the only doubly-occupied orbital correlated\footnote{Note that the presence of at
least one doubly-occupied orbital is important in order to have contributions from all possible
excitation classes in a second-order perturbation theory approach, allowing for a fair
comparison with the MRCISD method.}.
No shift was used in any CASPT2 calculation: neither real nor imaginary, nor IPEA.
In both regions where the avoided crossings happen at CASSCF level, we note a significant,
unphysical distortion of the MS-CASPT2 curves, but not for the other two methods.
The 1-2 AC is responsible for a ``hump'' in both the ground and first excited states,
while around the 2-3 AC we observe a clear artifact for the $1^1\Sigma^+$ state and, again, a
small hump on the $3^1\Sigma^+$ curve.
Remarkably, besides the issues in the AC regions, the MS-CASPT2 PECs fall right on top of the
MRCISD ones: this is not the case for a 2-state calculation, in which the three methodologies provide
three distinct results.
Around the equilibrium distance MS-CASPT2 is in very good agreement with MRCISD, with transition
energies to the first and second excited states underestimated by only 0.05 and 0.11 eV, respectively.
In contrast, XMS-CASPT2 overestimates these excitations by 0.2 eV and 0.25 eV, respectively.
On the other hand, the plot shown in \Cref{fig:LiF_reference} demonstrates the effectiveness
of XMS-CASPT2 in correcting the erratic behavior of the original theory, with PECs that are smooth
throughout the entire range of $r$.

\begin{figure}
  \centering
  \includegraphics{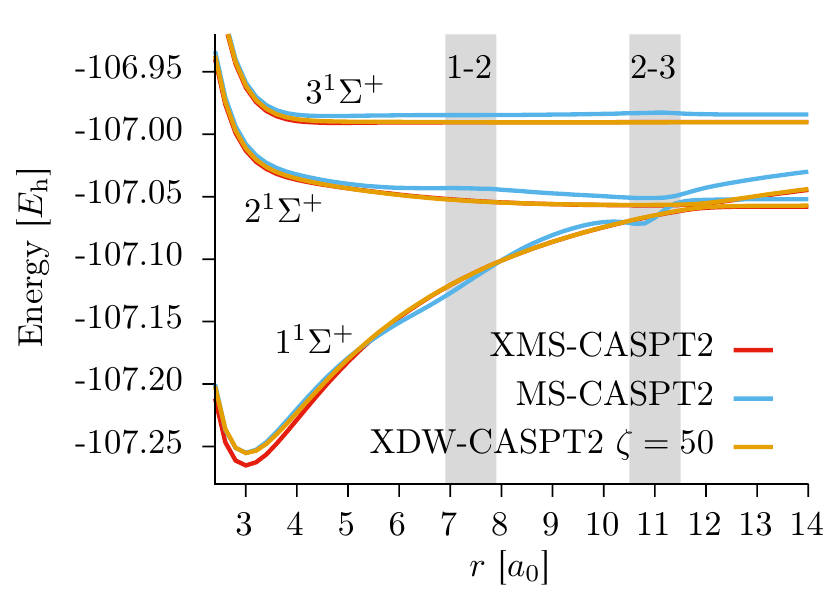}
  \caption{Potential energy curves of the three lowest $^1\Sigma^+$ states of lithium fluoride.
  Note that to a large extent the XMS-CASPT2 curves are covered by the XDW-CASPT2 ones.}
  \label{fig:LiF_z50}
\end{figure}

\begin{figure}
  \centering
  \includegraphics{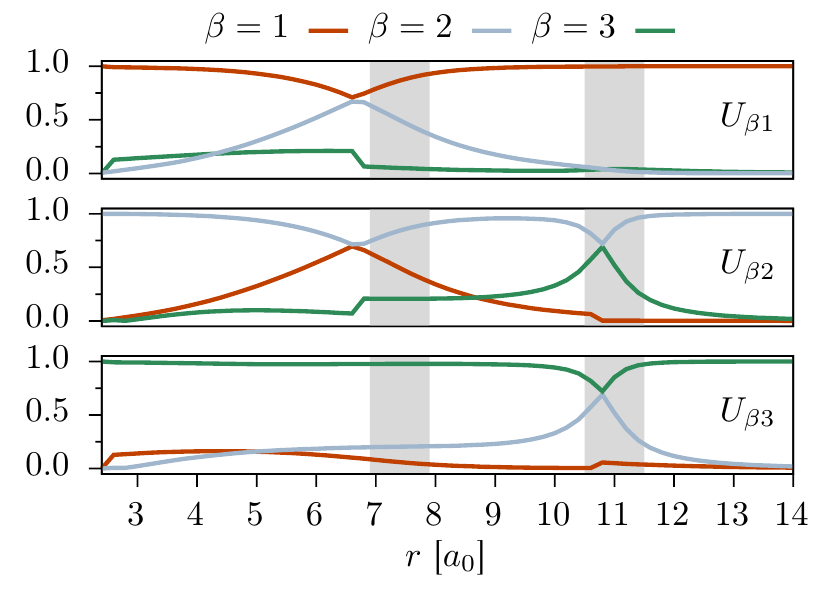}
  \caption{Absolute value of the elements $U_{\beta\alpha}$ of the rotation matrix mixing the
  zeroth-order CASSCF wave functions. The ground state ($\alpha=1$) is shown at the top, the
  first excited state ($\alpha=2$) in the center and the second excited state ($\alpha=3$) at
  the bottom. The zones highlighted in gray correspond to SA-CASSCF ACs.}
  \label{fig:LiF_XMS_evecs}
\end{figure}

Let us now investigate the performance of XDW-CASPT2 and study the dissociation of LiF as a
function of the exponent $\zeta$. Recalling that for $\zeta = 0$ all states receive the same
weight irrespective of their energy difference (and thus the methodology is exactly equivalent to
XMS-CASPT2), we show in \Cref{fig:LiF_z50} the results obtained by setting $\zeta = 50$.
The XDW-CASPT2 potential energy curves substantially overlap the XMS-CASPT2 ones for most of the
dissociation, showing no sign of artifacts at any place.
Crucially, the $1^1\Sigma^+$ state, and to some extent the $2^1\Sigma^+$ and $3^1\Sigma^+$ states,
smoothly slide over to the MS-CASPT2 curves for $r < 5$ $a_0$, with an excellent
agreement around the equilibrium distance.
To rationalize this result, we first analyze the structure of the transformation matrix that
diagonalizes $\hf^{sa}$ and inspect the magnitude of zeroth-order mixing among the states.
In \Cref{fig:LiF_XMS_evecs} we show the absolute value of the rotation matrix elements
$U_{\beta\alpha}$ (cf. \Cref{eq:Psi_0_rotated}) as a function of the internuclear distance.
The top plot represents the components of the ground state wave function. The magnitude of $U_{31}$,
i.e. the curve corresponding to $\beta=3$, never exceeds 0.25 for the entire range of distances,
meaning that the contribution of $\Pz_3$ to $\Pzt_{1}$ is very limited.
In contrast, the magnitude of $U_{21}$, i.e. the curve corresponding to $\beta=2$, increases when
approaching $r \approx 6.75$ $a_0$, with a peak in the vicinity of the SA-CASSCF avoided crossing.
Reciprocally, $U_{11}$ decreases in the same region substantially attaining the same value of
$U_{21}$ at $r \approx 6.75$ $a_0$, implying an equal mix of these two states.
Lastly, note that $U_{11}$ is approximately 1 for the most part of the plot, that is the
off-diagonal elements of $\hf^{sa}$ are very small and therefore $\Pzt_1 \approx \Pz_1$.
An analogous analysis for the other two plots leads to the following general observations.
The magnitude of mixing is a signature of the quasidegeneracies between the states:
around 6.75 $a_0$ the ground and first excited states are equally mixed, while the first and second
excited states mix just before 11 $a_0$. Both cases are around the ACs.
At $r= 3$ $a_0$, $U_{11} \approx U_{22} \approx U_{33} \approx 1$, meaning that the original CASSCF
wave functions are barely coupled by $\hf^{sa}$ and thus remain virtually the same after
the transformation.

\begin{figure}
  \centering
  \includegraphics{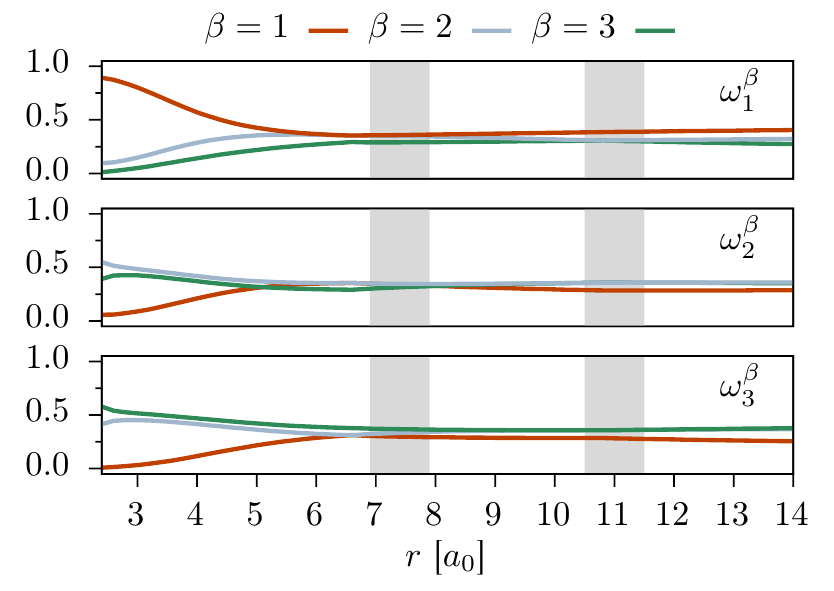}
  \caption{Weights $\omega_{\alpha}^{\beta}$ for $\zeta=50$. The ground state ($\alpha=1$) is
  shown at the top, the first excited state ($\alpha=2$) in the center and the second excited
  state ($\alpha=3$) at the bottom.}
  \label{fig:LiF_z50_w}
\end{figure}

To further understand the results shown in \Cref{fig:LiF_z50}, the weights used in the construction
of the density matrices are depicted in \Cref{fig:LiF_z50_w} in a plot similar to the one
for $U_{\beta\alpha}$.
At $r = 3$ $a_0$, the ground state weight $\omega_1^{\beta}$ with $\beta = 1$ is about 0.80, meaning
that $\mbf{\bar{D}}^1$ closely resembles $\mbf{D}^1$, thereby resulting in a Fock operator similar
to the MS-CASPT2. The latter is ultimately responsible for the very good agreement between the
XDW-CASPT2 and the MS-CASPT2 energy.
The densities of the other two states are instead approximately a 50\% mixture (central and bottom
plots); as a consequence, the energy of the $2^1\Sigma^+$ and $3^1\Sigma^+$ states is somewhere
in-between the MS-CASPT2 and XMS-CASPT2 one.
Note that such straightforward analogies are facilitated by the fact that the zeroth-order states
are very weakly coupled through $\hf^{sa}$ at $r=3$ $a_0$. In case of strong mixing, such an analysis
would be much harder.
At geometries with $r > 6$ $a_0$, the weights are roughly equal for all the states. This results
in Fock operators $\hfb^{\alpha}$ resembling $\hf^{sa}$ for $\alpha=1,2,3$ and thus XDW-CASPT2
essentially performs as XMS-CASPT2.

\begin{figure}
  \centering
  \includegraphics{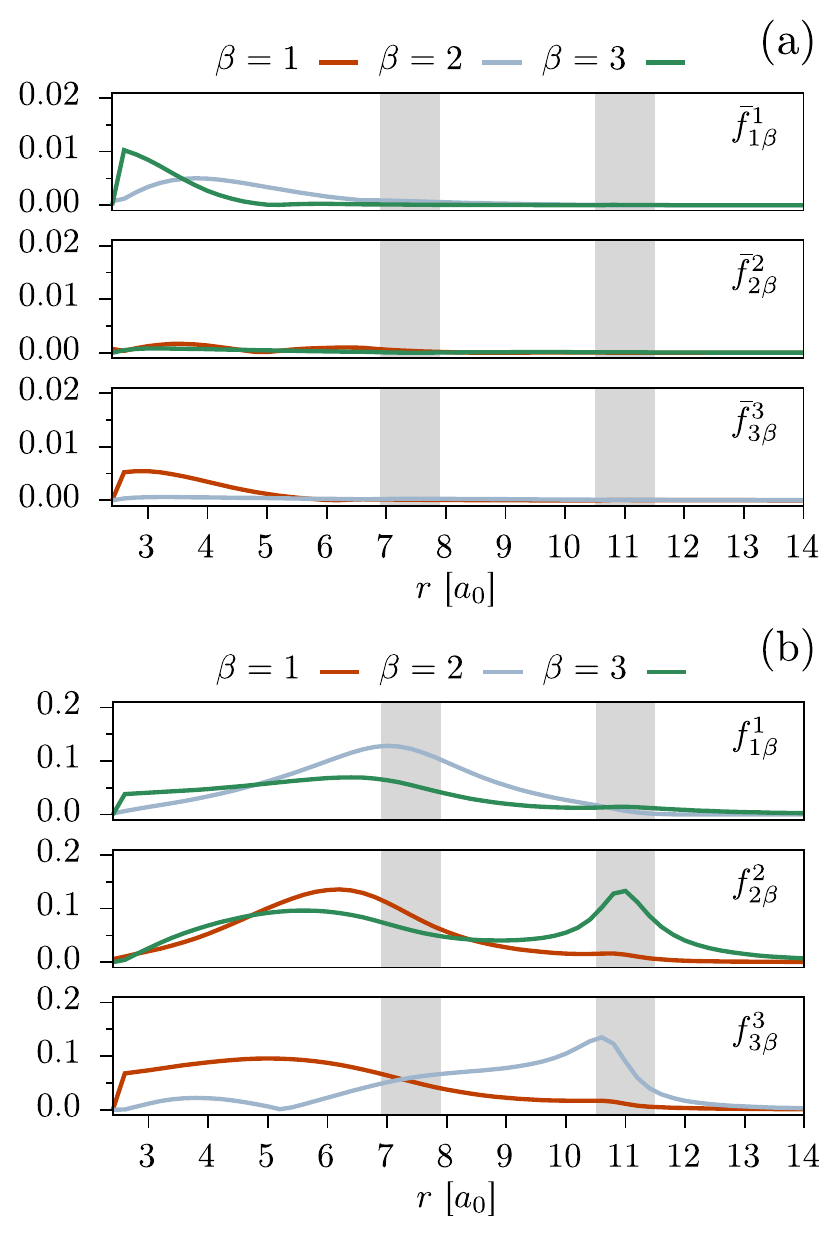}
  \caption{Absolute values of the Fock operator off-diagonal entries for
  (a) XDW-CASPT2 with $\zeta=50$ (elements $\bar{f}^{\gamma}_{\alpha\beta}$) and
  (b) MS-CASPT2 (elements $f^{\gamma}_{\alpha\beta}$).
  For each method (three plots), the ground state ($\alpha=1$) is shown at the top, the
  first excited state ($\alpha=2$) in the center and the second excited state ($\alpha=3$)
  at the bottom.
  Note that the Fock operator used to compute the couplings is different for each state and
  only the case $\gamma=\alpha$ is of relevance.}
  \label{fig:LiF_z50_MS_f}
\end{figure}

The invariance properties of XDW-CASPT2 rely on the assumption made in \Cref{eq:fab_XDW_approx_0},
i.e. that the off-diagonal terms $\bar{f}^{\gamma}_{\alpha\beta} = \braket{\Pzta|\hfb^{\gamma}|\Pztb}$
are approximately zero. It is interesting to investigate if this is the case for LiF.
In \Cref{fig:LiF_z50_MS_f} (a) we show the absolute values of the Fock couplings for each of the
three states.
The largest elements are observed for the ground state around the equilibrium distance; this is not
surprising since the Fock operator is essentially state-specific in that region.
On the other hand the opposite is true past $r = 6$ $a_0$, with the three Fock operators being
roughly equivalent and equal to $\hf^{sa}$ (cf. \Cref{fig:LiF_z50_w}).
Recalling that the rotated zeroth-order states diagonalize $\hf^{sa}$, their coupling must be
approximately zero.
The elements $\bar{f}_{\alpha\beta}^{\alpha}$, albeit different from zero, are in practice small
enough to yield smooth potential energy curves.
As a matter of comparison, the MS-CASPT2 zeroth-order off-diagonal elements between the original
CASSCF states are shown in \Cref{fig:LiF_z50_MS_f} (b): the difference is striking, with values
that are one order of magnitude larger compared to XDW-CASPT2.
The strongest couplings are around the avoided crossings, exactly where MS-CASPT2 performs poorly.

\begin{figure}
  \centering
  \includegraphics{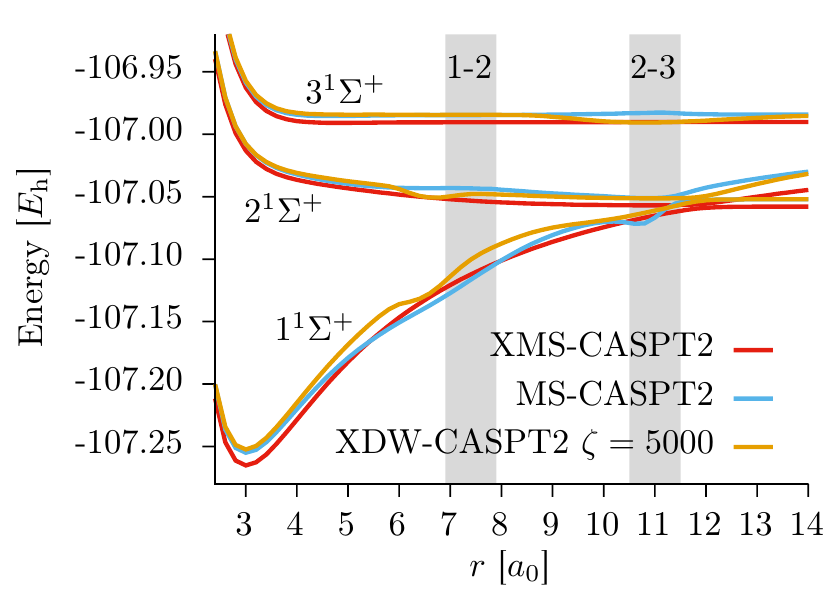}
  \caption{Potential energy curves of the three lowest $^1\Sigma^+$ states of lithium fluoride.}
  \label{fig:LiF_z5000}
\end{figure}

\begin{figure}
  \centering
  \includegraphics{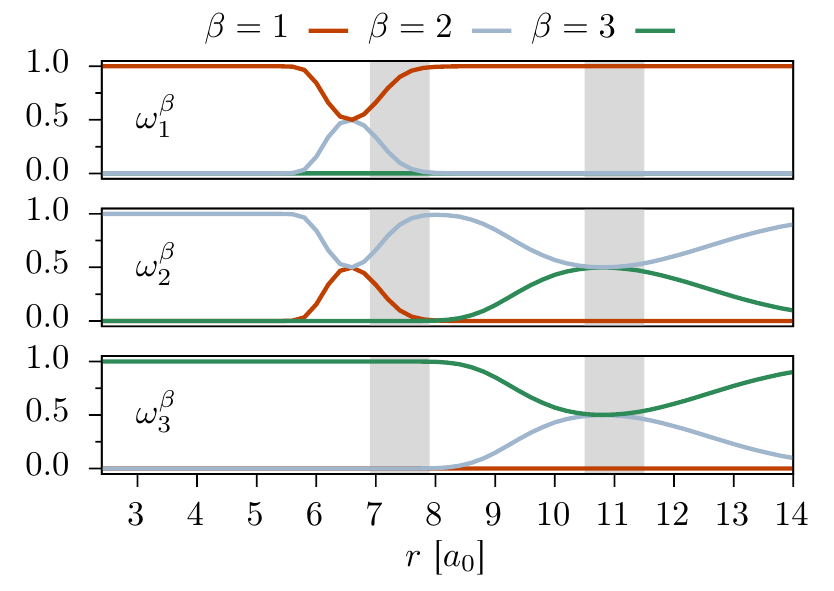}
  \caption{Weights $\omega_{\alpha}^{\beta}$ for $\zeta=5000$.}
  \label{fig:LiF_z5000_w}
\end{figure}

\begin{figure}
  \centering
  \includegraphics{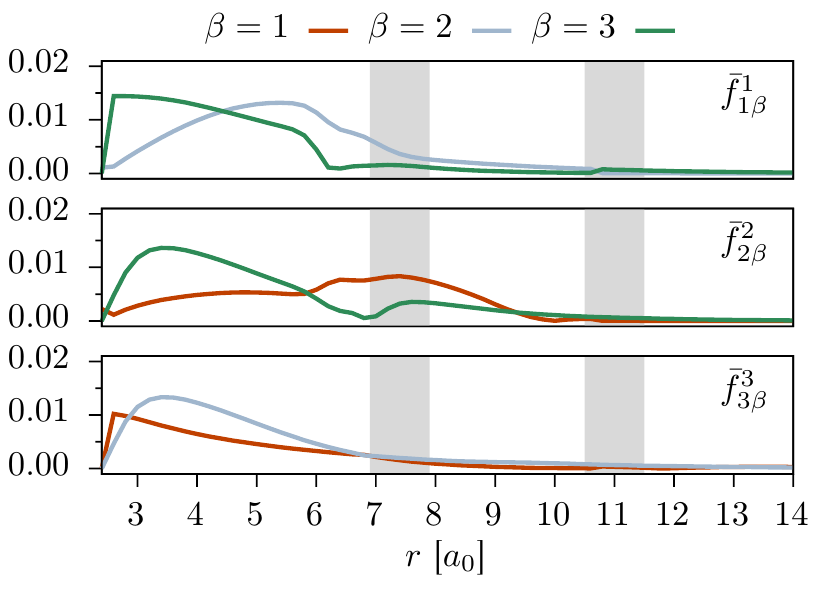}
  \caption{Absolute values of the elements $\bar{f}^{\alpha}_{\alpha\beta}$ for $\zeta=5000$.}
  \label{fig:LiF_z5000_f}
\end{figure}

\begin{figure}
  \centering
  \includegraphics{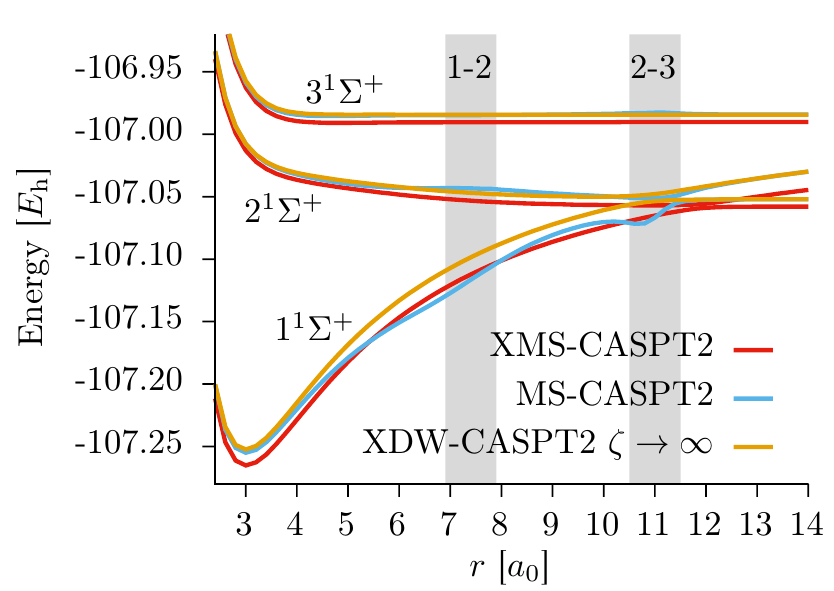}
  \caption{Potential energy curves of the three lowest $^1\Sigma^+$ states of lithium fluoride.}
  \label{fig:LiF_zInf}
\end{figure}

Increasing the value of $\zeta$ sharpens the transition between state-specific and state-average
regimes. As already observed in the context of DW-DSRG\cite{Li2019a}, this leads to the
appearance of wiggles along the potential energy curves due to sudden changes of the
zeroth-order weights.
This behavior can be seen in \Cref{fig:LiF_z5000} for $\zeta = 5000$.
For instance, near the 1-2 AC, the XDW-CASPT2 curve for state $2^1\Sigma^+$ rapidly switches
between the XMS-CASPT2 and MS-CASPT2 references.
Inspection of the weights in \Cref{fig:LiF_z5000_w} reveals a clear correlation between the weights
$\omega_{\alpha}^{\beta}$ and these oscillations. Whenever the weights undergo a rapid and
significant change, the energy does so accordingly.
Despite this oscillatory behavior, the off-diagonal elements of the Fock operator for $\zeta=5000$
are in the same order of magnitude as for $\zeta=50$, hence still 10-fold less than MS-CASPT2,
as can be seen in \Cref{fig:LiF_z5000_f}. Therefore, it appears that the cause of the wiggles in the
PECs is \emph{not} due to the diagonal approximation.
\newline
Lastly, the results obtained taking the limit $\zeta \to \infty$ are reported in \Cref{fig:LiF_zInf}.
For this case the weights never change and correspond to unit vectors, hence the densities are
state-specific: $\mbf{\bar{D}}^{\alpha} = \mbf{\tilde{D}}^{\alpha}$.
This leads to potential energy curves that mostly overlap with the MS-CASPT2 ones, however
\emph{without} the artifacts around the SA-CASSCF near-degeneracies.
Notably, state-specific Fock operators built with densities $\mbf{\tilde{D}}^{\alpha}$ do not
couple the states as strong as the original operators, $\hf^{\alpha}$, since
the zeroth-order off-diagonal elements for $\zeta \to \infty$ are as large as those for
$\zeta=5000$ (see Supporting Information).
This result is important because it corroborates the conjecture that the PEC wiggles observed for
intermediate values of $\zeta$ are strictly caused by the rapid change of the weights.

\subsection{Conical intersection in allene}

\begin{figure}
  \centering
  \includegraphics[width=8cm]{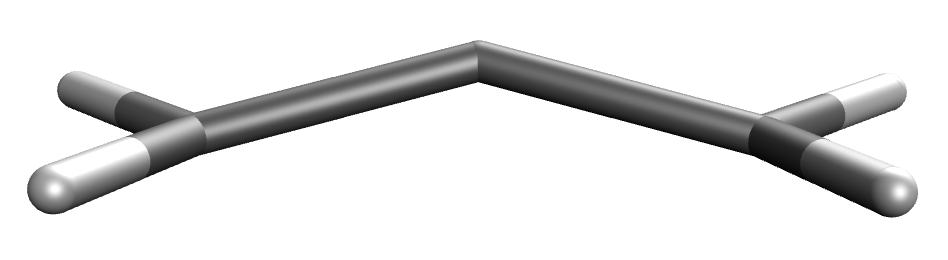}
  \caption{The $1^1$A$'$ and $2^1$A$'$ MECI geometry of the allene molecule.}
  \label{fig:allene}
\end{figure}

Projection of the zeroth-order Hamiltonian onto the individual states of the model space defines
a MRPT that is \emph{not} invariant under unitary transformation of the model states.
Failure to satisfy \Cref{eq:fab_XDW_approx_0} leads to unphysical results at conical
intersections or in the vicinity of avoided crossings.
This situation has been already observed in the LiF dissociation, however a more challenging
test is that of the minimum energy conical intersection (MECI) of the distorted
allene molecule, depicted in \Cref{fig:allene}.
Around that point, the $1^1$A$'$ and $2^1$A$'$ states are quasidegenerate and thus only the space
spanned by them is well-defined.
In order to investigate the behavior of the various CASPT2 variants in this situation, we performed
two-dimensional, non-relaxed scans by varying the C-C-C bend angle and the C-C-C-H torsional angle
in steps of 0.25 degrees in the range of $-10$ to $+10$ degrees from the CASSCF
MECI point, respectively\footnote{Note that the C-C-C-H torsion angle is simultaneously changed on
both sides of the molecule in order to preserve the C$_s$ symmetry.}.
The computational details are the same as in Ref. \citenum{Granovsky2011}, and are fully described
in the Supporting Information; here we report only the essential points.
The reference wave functions were obtained in a SA-CASSCF calculation with 4 electrons in 3
orbitals of a$'$ symmetry and 1 orbital of a$''$ symmetry.
This amounts to a complete active space of 12 totally symmetric configuration state functions,
thus allowing to study the behavior of the potential energy surface as a function of the number of
states, up to the complete active space limit.
Given that the calculation focuses on the $1^1$A$'$ and $2^1$A$'$ conical intersection, the CASSCF
orbital optimization was carried out for the two lowest states only, while the remaining 10 states
were obtained by diagonalization of the configuration interaction matrix.
The basis set used was the \textsc{gamess (us)}-style variation of the Dunning-Hays basis, augmented
by a single polarization spherical $d$ function on each carbon.

\begin{figure*}
  \centering
  \includegraphics{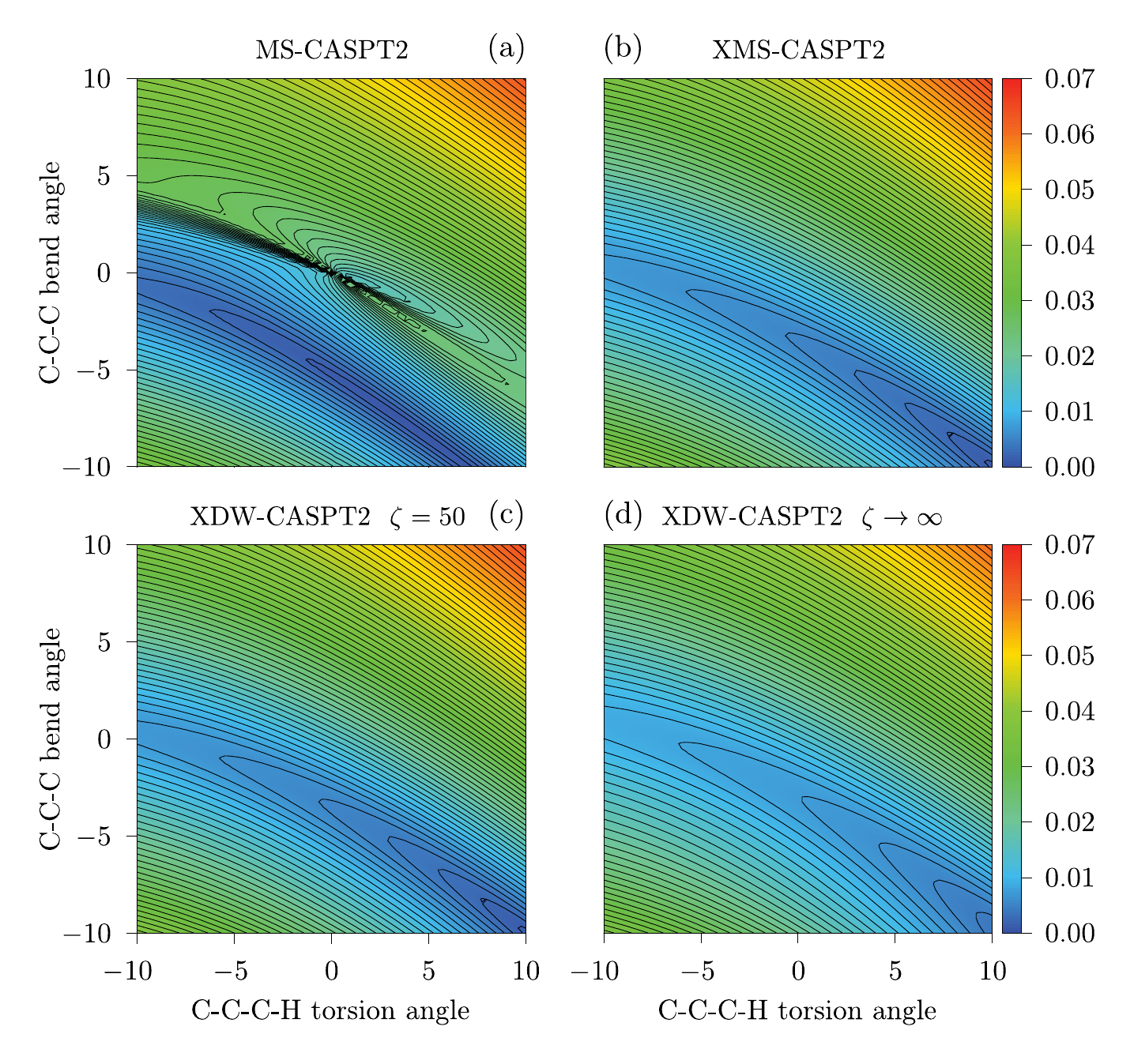}
  \caption{Colormapped isosurface plot of the absolute energy difference (in $\Eh$) between
  the $1^1$A$'$ and $2^1$A$'$ states for a model space including 2 states.
  The same calculation was carried out with different methodologies: (a) MS-CASPT2,
  (b) XMS-CASPT2, (c) XDW-CASPT2 with $\zeta=50$ and (d) XDW-CASPT2 with $\zeta\to\infty$.}
  \label{fig:allene_2x2}
\end{figure*}

In \Cref{fig:allene_2x2} we report the colormapped isosurface plots of the energy difference between
the $1^1$A$'$ and $2^1$A$'$ states computed with a model space spanned by the 2 lowest roots only.
\Cref{fig:allene_2x2} (a) shows the result obtained with MS-CASPT2.
At the CASSCF MECI point --- the origin of the plot --- there is a singularity and the surface
around this point is completely compromised, showing the deficiency of this methodology.
On the contrary, as can be seen from \Cref{fig:allene_2x2} (b), the surface obtained with XMS-CASPT2
does not show any sign of artifacts, demonstrating the importance of invariance to obtain physically
sound results.
The plot in \Cref{fig:allene_2x2} (c) illustrates the behavior of XDW-CASPT2 for $\zeta=50$.
The surface is virtually identical to that obtained with XMS-CASPT2, a result that is easily explained
upon analyzing the density weights. Recalling that the model space only has a dimension of two, the
largest and smallest values of $\omega_{\alpha}^{\beta}$ observed in the entire scan were 0.53 and
0.47, respectively, meaning that the difference between the XDW-CASPT2 and XMS-CASPT2 partitions
are very small across the board.
Lastly, in \Cref{fig:allene_2x2} (d) is shown the surface obtained by letting $\zeta \to \infty$.
Remarkably, albeit the use of purely state-specific operators, the PES around the MECI point is
perfectly smooth. The overall morphology is analogous to the last two cases, even though
a slightly larger width of the potential well is noticeable.

\begin{figure}
  \centering
  \includegraphics{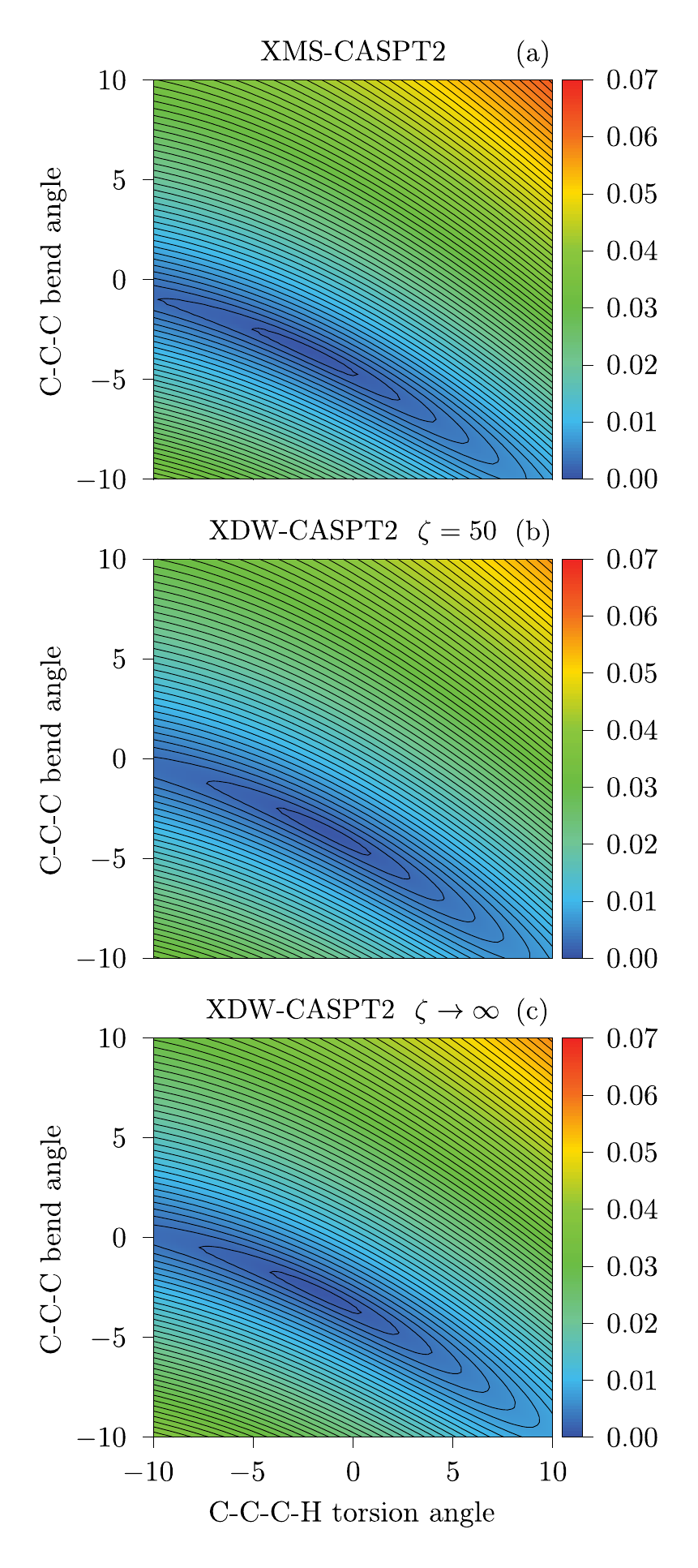}
  \caption{Colormapped isosurface plot of the absolute energy difference between the
  $1^1$A$'$ and $2^1$A$'$ states for a 12-state model space computed with different
  methodologies: (a) XMS-CASPT2, (c) XDW-CASPT2($\zeta=50$) and
  (c) XDW-CASPT2($\zeta\to\infty$).}
  \label{fig:allene_12x12}
\end{figure}

Repeating the same calculation with XMS-CASPT2 including all 12 states of the complete
active space results again in a smooth surface as shown in \Cref{fig:allene_12x12} (a).
The position of the MECI at the correlated level changes according to the number of states,
and for XMS-CASPT2 is substantially converged with a model space of 6 states (see Supporting
Information).
In \Cref{fig:allene_12x12} (b) and (c) we report the result obtained with XDW-CASPT2 for $\zeta = 50$
and $\zeta\to\infty$.
Once again the PESs are smooth everywhere and on par with the XMS-CASPT2 one.
Moreover, we note that all three plots of \Cref{fig:allene_12x12} are remarkably similar
to the one obtained by \citet{Granovsky2011} with extended MCQDPT.
In contrast to the 2-state case, the dynamical weights obtained with $\zeta = 50$ are significantly
different from the state-average ones.
The Fock operators $\hfb^1$ and $\hfb^2$ are substantially defined by the first four states,
since the weights assigned starting from the fifth one are less than 0.03, thereby contributing
little to nothing to the $1^1$A$'$ and $2^1$A$'$ 1-RDMs.
Crucially, this does not imply that PESs obtained with a model space of dimension four are the same
as those obtained with one of higher dimension.
Both \Cref{eq:Psi_0_rotated} and \Cref{eq:boltzmann_weight} directly depend on the total number of
model states and their wave function.
As a result, different model space dimensions give rise to distinct partitionings of $\hH$,
which are ultimately coupled together in a non-trivial way through the formation of the second-order
effective Hamiltonian.

\subsection{Vertical excitation energies}

One of the design objectives of XDW-CASPT2 is to maintain the accuracy of MS-CASPT2 in the
calculation of transition energies.
If we consider a molecule in its electronic ground state, the dynamical weighting scheme is
such that when the energy gap to the first excited state is larger than $\zeta^{-1/2}$, then the
density matrices of both these states will barely mix with each other, remaining predominantly
state-specific. It is reasonable to assume that in this situation the energy separation between
these states is sizable and that their associated wave functions have well-defined, but distinct
character. Hence, the rotated reference states obtained from \Cref{eq:Psi_0_rotated} will be similar
to the original ones: $\Pzta \approx \Pza$.
Under these circumstances, all the quantities in XDW-CASPT2 will not be very different
from those in MS-CASPT2, such that we expect the two methods to have a comparable accuracy.
This was indeed observed in the previous section for lithium fluoride.
In principle, the same logic applies when the model space dimension is larger than two:
as long as all states are energetically well separated from each other and the rotated reference
wave functions maintain their original character, we expect similar results for XDW-CASPT2 and
MS-CASPT2.
A different and much more complicated situation occurs when many model states lie within a limited
region of the spectrum and interact strongly with each other at zeroth-order. Although it is
conceptually easy to visualize the amount of density mixing by inspecting the weights
$\omega_{\alpha}^{\beta}$, the fact that the rotated model states are linear combinations of the
original reference wave functions, makes it hard to rationalize the physical content of the Fock
operator in these terms.
The three different cases are summarized in Scheme \ref{fig:transitions_scheme}.
\begin{scheme*}
  \centering
  \includegraphics{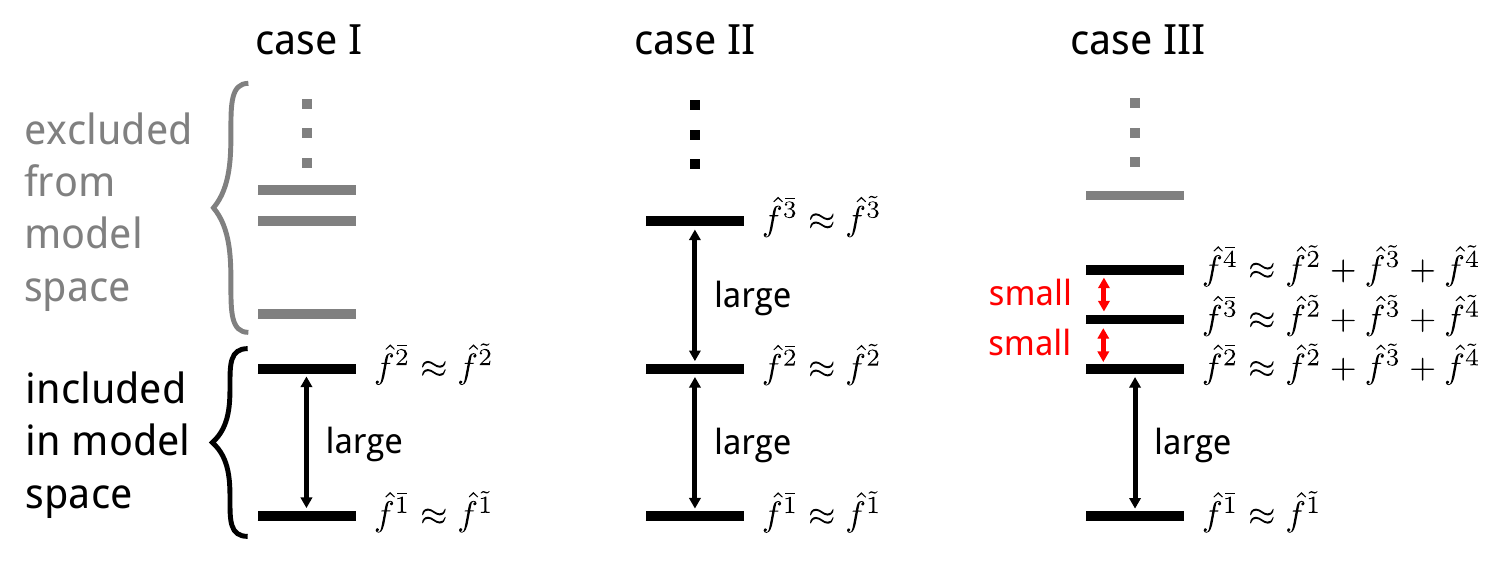}
  \caption{Three main scenarios for the calculation of excited states energies.
  In case I only the well-separated ground and first excited states are included
  in the model space. In case II many states are included in the calculation, but all
  of them are well separated. In case III several low-lying excited states are
  included in the model space and these are energetically very close to each other.
  Therefore, their Fock operators will be approximately state-average in contrast to
  the other cases.}
  \label{fig:transitions_scheme}
\end{scheme*}

\begin{figure*}
  \centering
  \includegraphics[width=\textwidth]{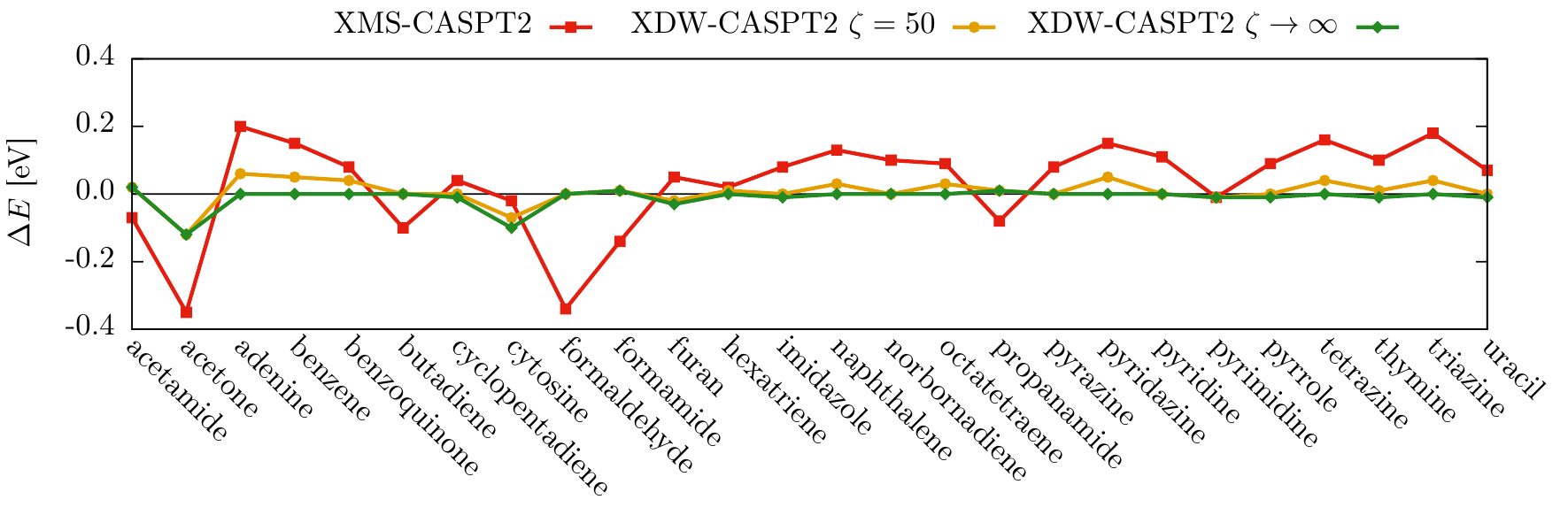}
  \caption{Signed deviations of singlet vertical excitation energies with respect to MS-CASPT2.}
  \label{fig:transitions_deviations}
\end{figure*}

\begin{figure}
  \centering
  \includegraphics{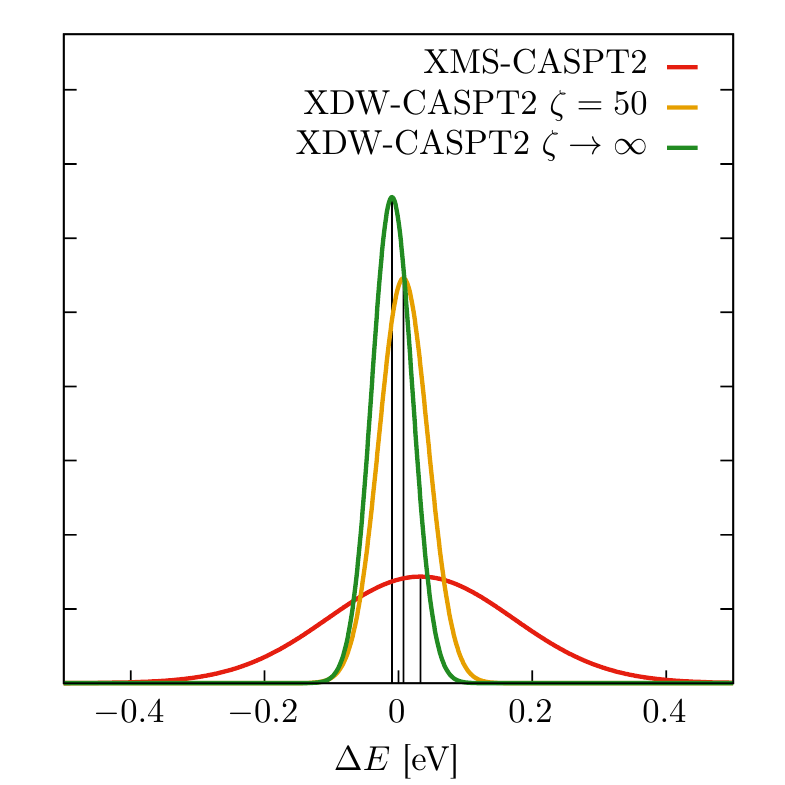}
  \caption{Normal distributions of excitation energy deviations with respect to MS-CASPT2.}
  \label{fig:transitions_gaussian}
\end{figure}

In order to assess the accuracy of XDW-CASPT2 for the calculation of electronically excited states,
we computed the vertical energy gap between the ground and the first excited singlet state for a
series of small to medium organic compounds and compared the results to MS-CASPT2.
This case corresponds to the first scenario illustrated in Scheme \ref{fig:transitions_scheme}.
The molecules were taken from Thiel's benchmark set\cite{Schreiber2008}, excluding ethene and
cyclopropene since no singlet excited state was considered for these two systems.
In order to appreciate the effects of the dynamical weighting scheme, the calculated first excited
states always belonged to the same irreducible representation as the ground state.
The geometries were taken from Ref. \citenum{Schreiber2008} and correspond to structures
optimized at MP2/6-31G* level of theory. The reference wave functions were obtained by a 2-state
SA-CASSCF calculation using the TZVP basis set\cite{Schafer1994a} and the RICD
approximation\cite{Aquilante2007}.
Full computational details are available in the Supporting Information.
Vertical transition energies were calculated with MS-CASPT2, XMS-CASPT2 and XDW-CASPT2 with two
values of $\zeta$ and setting the IPEA shift to zero for all methods. A real shift was used when
necessary and equally applied to all methods in order to obtain comparable energies.\\
In \Cref{fig:transitions_deviations} is reported the deviation of XMS-CASPT2 and XDW-CASPT2 vertical
excitation energies with respect to MS-CASPT2, which is used as the reference since the design
objective is to reproduce its value.
The largest deviation is observed for XMS-CASPT2, with a general tendency to slightly overestimate
the excitation energies by up to 0.2 eV. In two cases, acetone and formaldehyde, the energy is instead
considerably underestimated: these transitions correspond to the largest of the entire set,
8.93 eV and 10.06 eV, respectively, and highlight the difficulty of the state-average Fock operator
to deal with states that energetically so far apart. Note that the MS-CASPT2 values,
9.28 eV and 10.40 eV, agree well with CC3/TZVP\cite{Schreiber2008}, with transitions estimated at
9.65 eV and 10.45 eV, such that the underestimation of XMS-CASPT2 appears to be a true deficiency
of the methodology.
The results of XDW-CASPT2 are instead on par with MS-CASPT2. The values for the transitions
are exactly reproduced for several systems, more so for $\zeta\to\infty$ than $\zeta=50$,
even though the general performance of both is virtually the same.
The accuracy of the three methodologies is well captured by normal distributions of the energy
deviations with respect to MS-CASPT2 as shown in \Cref{fig:transitions_gaussian}.
Note that we do not want to make any claim that the deviations are normally distributed, but just
point out that this plot neatly summarizes the results shown in \Cref{fig:transitions_deviations}.
Despite the mean of all three methods is quite close to the MS-CASPT2 one, XDW-CASPT2 is clearly a
more reliable choice than XMS-CASPT2, at least for the scenario considered here.
In terms of mean absolute deviations, XMS-CASPT2 excitation energies differ by 0.12 eV on average,
whereas the agreement is excellent for XDW-CASPT2, with a discrepancy of only 0.02 eV and 0.01 eV
for $\zeta=50$ and $\zeta\to\infty$, respectively.

\section{Conclusions}

In this work we have proposed and investigated a new variant of the CASPT2 method. By a careful
analysis of the properties of MS-CASPT2 and XMS-CASPT2, we have identified the two key components
that characterize the success of each variant and included them in the newly developed XDW-CASPT2
approach.
First, diagonalization of the state-average Fock operator in the reference basis provides
a new set of zeroth-order states. Second, this is followed by the construction of state-specific Fock
operators with dynamically adjusted weights that depend on the energy separation between the states.
These operators are then used to partition the Hamiltonian in a MS-CASPT2 calculation.
The resulting method is approximately invariant under unitary transformations of the model states, a
property that ensures a physical behavior in the vicinity of avoided crossings and conical
intersections, and at the same time shows an accuracy comparable to conventional MS-CASPT2.
The dynamical weighting scheme introduces a parameter $\zeta$ which acts as a threshold controlling
the state-specificity of the Fock operator, thereby allowing the method to interpolate between
XMS-CASPT2 and MS-CASPT2 (with rotated reference functions).
Importantly, even though XDW-CASPT2 employs the diagonal approximation, in practice it approximately
satisfies all important properties listed by \citet{Granovsky2011}.

The reliability of XDW-CASPT2 is demonstrated in the typical benchmark system LiF, whose
avoided crossings represent a difficult task for multireference approaches.
The obtained potential energy curves overlap with the XMS-CASPT2 ones in the regions where the
underlying zeroth-order states are quasidegenerate, hence do not show the wiggles typical of
MS-CASPT2, but at the same time the vertical transitions to the first two excited states are in
better agreement than XMS-CASPT2 with the reference MRCISD values.
The robustness of XDW-CASPT2 is further tested by studying the conical intersection in the allene
molecule, for which smooth PESs where obtained for different values of $\zeta$ and dimensions
of the model space.
At last, vertical excitation energies are shown to be in almost perfect agreement with MS-CASPT2 for
singlet transitions in a set of 26 organic compounds, unlike XMS-CASPT2 that shows an average
deviation in the order of 0.1 eV and maximum deviations as large as 0.4 eV.

The XDW-CASPT2 method can be viewed as a bridge between MS-CASPT2 and XMS-CASPT2, drawing the best
from both these approaches and providing a valid alternative to other quasidegenerate multireference
perturbation theories.
Moreover, being based on the CASPT formalism constitutes a practical advantage: any
existing implementation can be easily adapted to provide XDW-CASPT2 as an option
and at the same time it only requires an additional input parameter from the final user.
The similarity with its parent theory also means that XDW-CASPT2 can also be used with zeroth-order
wave functions obtained with modern approaches such as the density matrix renormalization
group\cite{White1992}.
From the computational perspective, the only difference with (X)MS-CASPT2 is a small overhead for
the construction of the dynamically weighted densities and thus it is applicable to the same kind
of systems where the parent methods are an option.
At last, we envision XDW-CASPT2 to be a very interesting method in the context of \textit{ab initio}
molecular dynamics, once the restriction of imposing molecular symmetries is lifted, e.g. through
the use of the off-diagonal elements of the full Hamiltonian.

\begin{acknowledgement}
  S.B. acknowledges the Swiss National Science Foundation (SNSF) for the funding received
  through the Early Postdoc.Mobility fellowship (grant number P2SKP2\_184034).
  R.L. acknowledges the Swedish Research Council (VR, grant number 2016-03398).
\end{acknowledgement}

\begin{suppinfo}
  In the Supporting Information we provide additional results on the dissociation of lithium
  fluoride and on the conical intersection of the distorted allene molecule and we report
  supplementary computational details on the allene molecule as well as information
  necessary to reproduce the calculation of the singlet vertical excitation energies.
\end{suppinfo}

\bibliography{references}

\end{document}


\clearpage
\section{Avoided Crossing in LiF}

\begin{figure}
  \centering
  \includegraphics{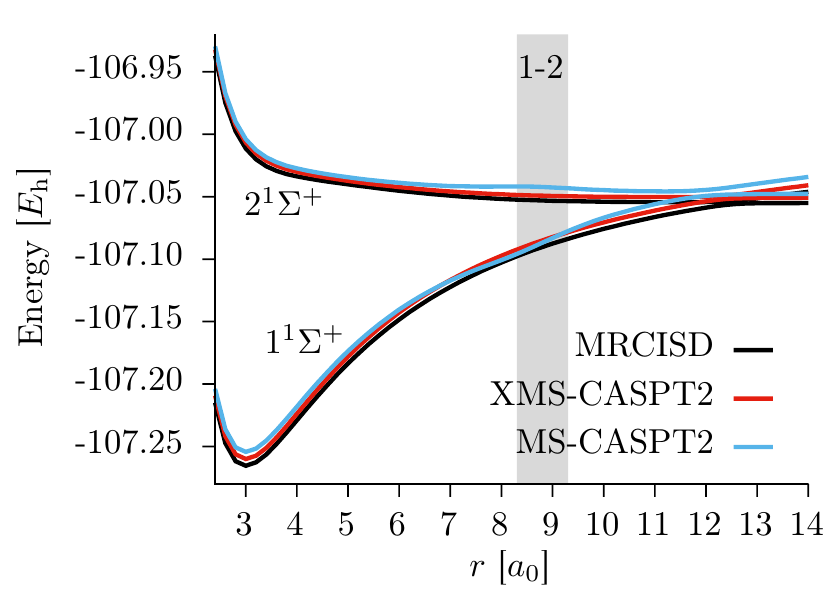}
  \caption{Potential energy curves of the two lowest $^1\Sigma^+$ states of lithium
  fluoride. The zones highlighted in gray correspond to the avoided crossing regions
  at the CASSCF level of theory.}
  \label{fig:LiF_reference_2states}
\end{figure}

\begin{figure}
  \centering
  \includegraphics{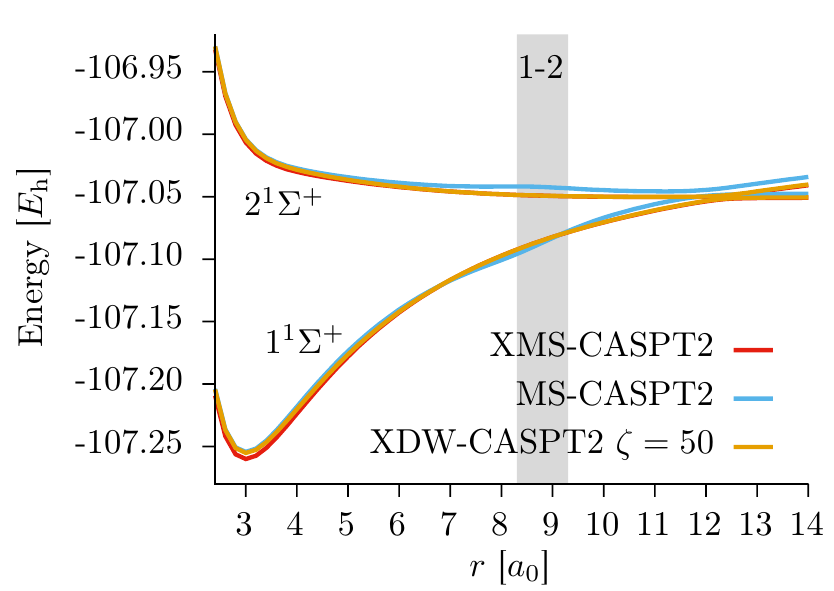}
  \caption{Potential energy curves of the two lowest $^1\Sigma^+$ states of lithium fluoride.
  Note that to a large extent the XMS-CASPT2 curves are covered by the XDW-CASPT2 ones.}
  \label{fig:LiF_2states_z50}
\end{figure}

\begin{figure}
  \centering
  \includegraphics{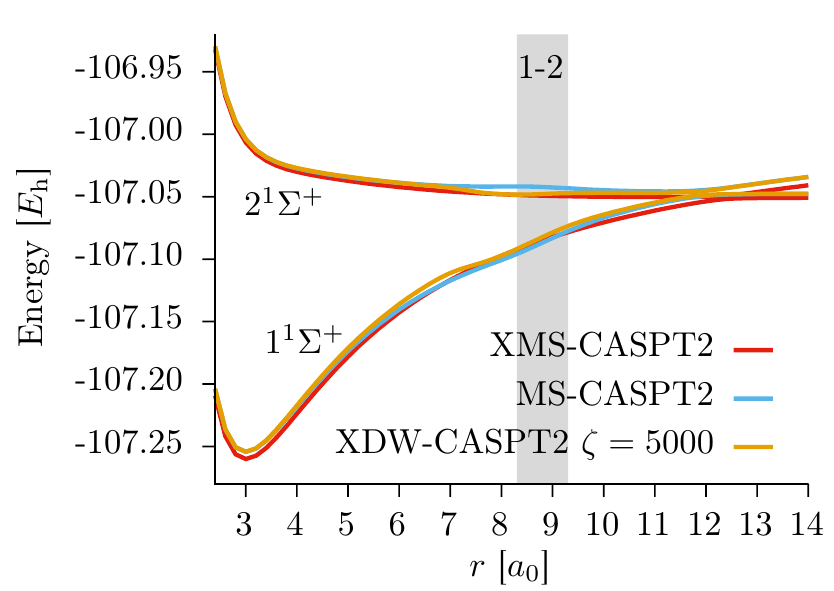}
  \caption{Potential energy curves of the two lowest $^1\Sigma^+$ states of lithium fluoride.}
  \label{fig:LiF_2states_z5000}
\end{figure}

\begin{figure}
  \centering
  \includegraphics{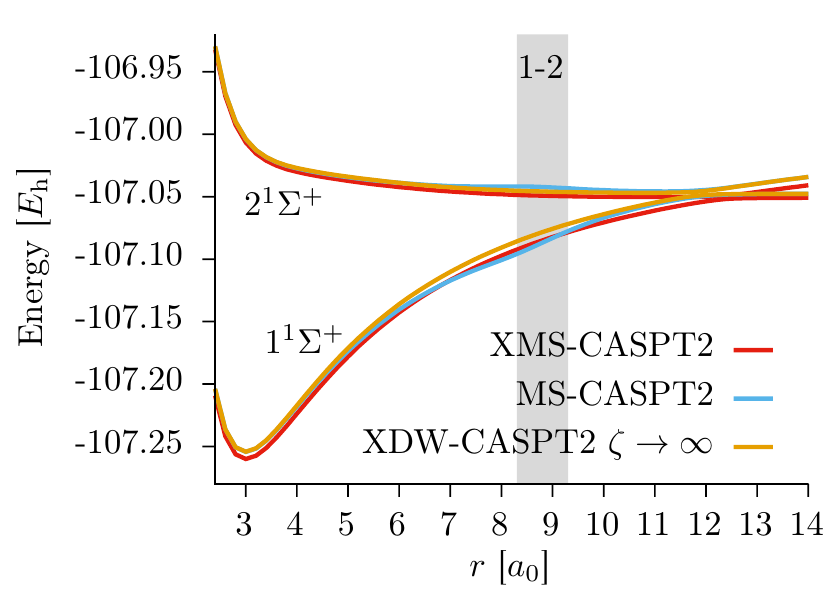}
  \caption{Potential energy curves of the two lowest $^1\Sigma^+$ states of lithium fluoride.}
  \label{fig:LiF_2states_zInf}
\end{figure}

\begin{figure}
  \centering
  \includegraphics{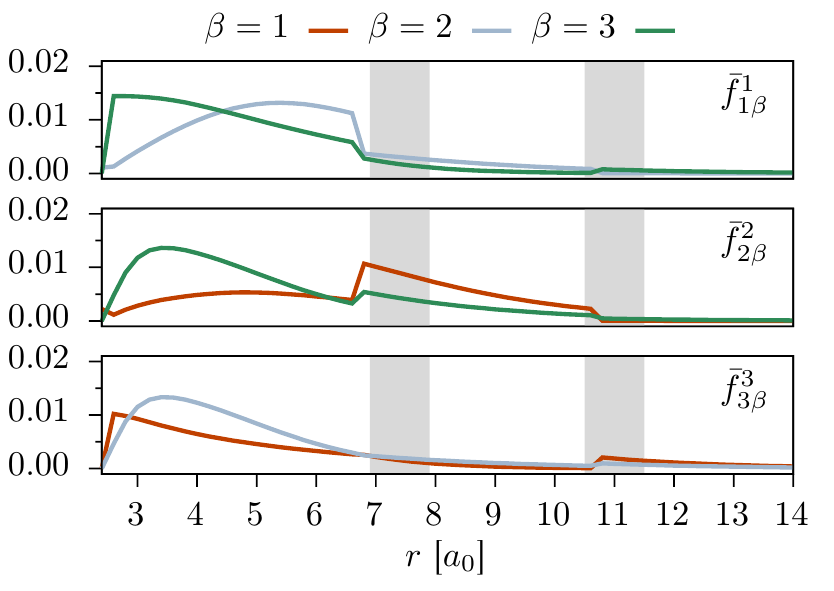}
  \caption{Absolute values of the elements $\bar{f}^{\alpha}_{\alpha\beta}$ for $\zeta\to\infty$.
  From top to bottom the couplings are for the ground, first excited and
  second excited state, respectively. Note that for each state (plot), the Fock
  operator used to compute the couplings is different.}
  \label{fig:LiF_zInf_f}
\end{figure}

\begin{figure}
  \centering
  \includegraphics{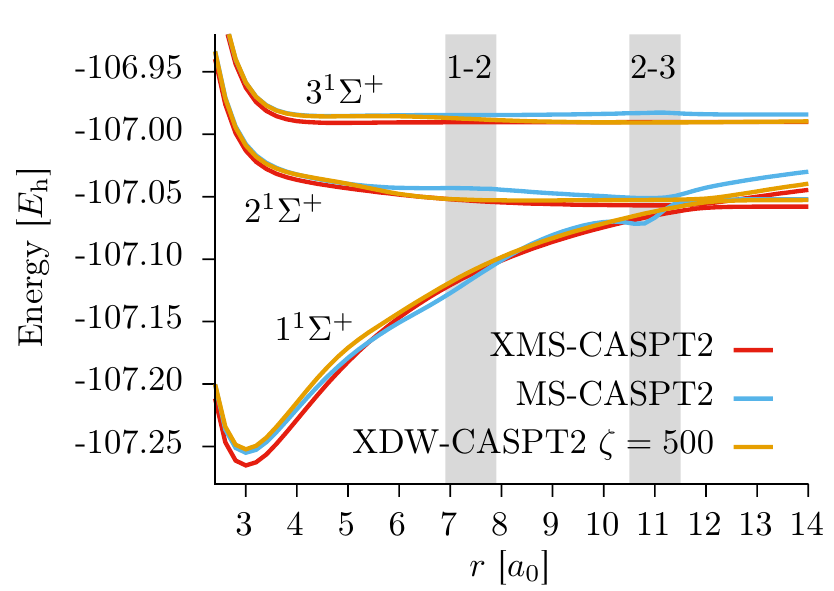}
  \caption{Potential energy surface of the three lowest $^1\Sigma^+$ states of lithium fluoride.}
  \label{fig:LiF_z500}
\end{figure}

\begin{figure}
  \centering
  \includegraphics{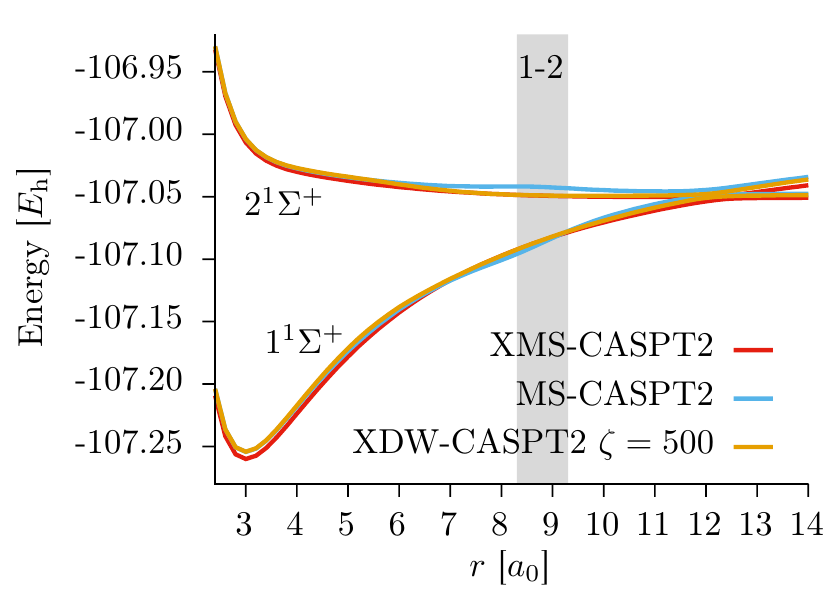}
  \caption{Potential energy curves of the two lowest $^1\Sigma^+$ states of lithium fluoride.}
  \label{fig:LiF_2states_z500}
\end{figure}

\begin{figure}
  \centering
  \includegraphics{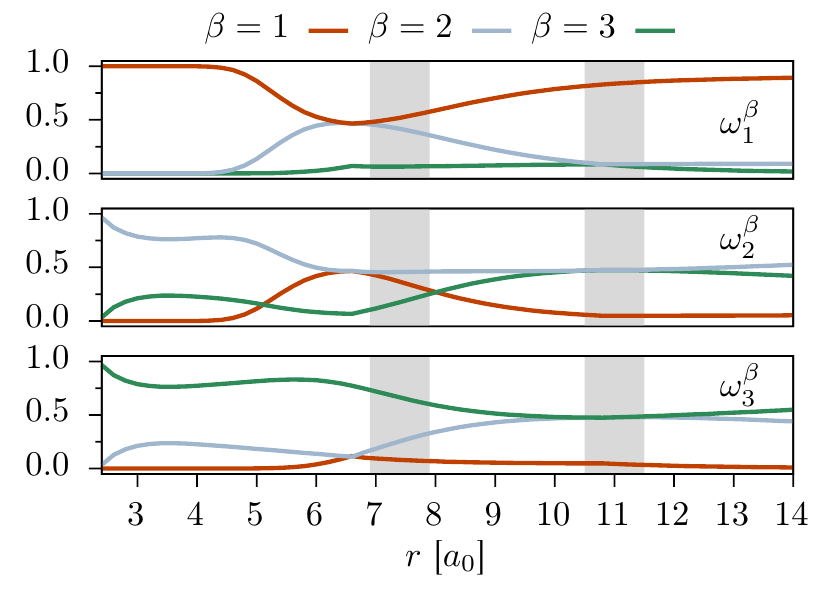}
  \caption{Weights $\omega_{\alpha}^{\beta}$ for $\zeta=500$.}
  \label{fig:LiF_z500_w}
\end{figure}

\begin{figure}
  \centering
  \includegraphics{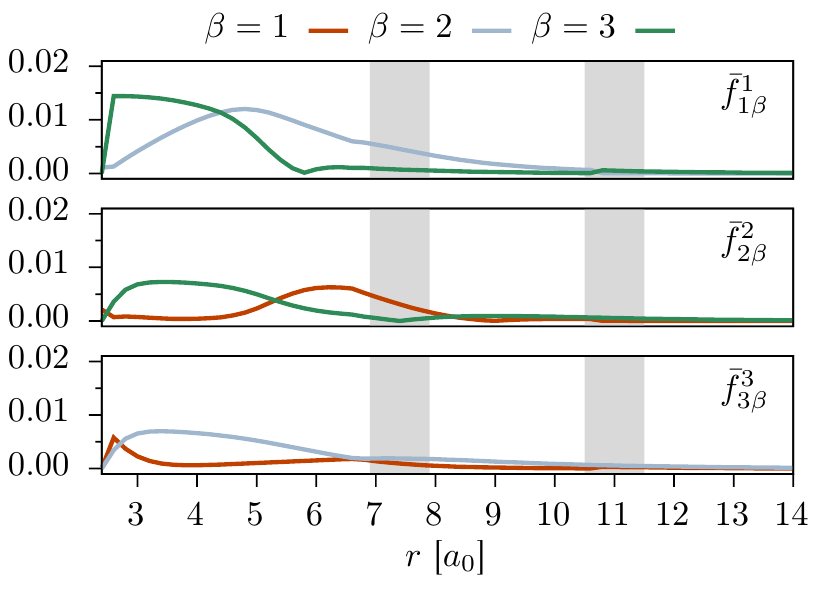}
  \caption{Absolute values of the elements $\bar{f}^{\alpha}_{\alpha\beta}$ for $\zeta=500$.}
  \label{fig:LiF_z500_f}
\end{figure}

In this section we show additional results for the dissociation of lithium fluoride.
First, we report the potential energy curves in the case only two states are included in the
calculation. The computational details are the same as reported in the main text, with the only
difference being the number of roots considered in both the SA-CASSCF and the various CASPT2
calculations.
In \Cref{fig:LiF_reference_2states} we report the results of the reference calculation with MRCISD.
The results for XDW-CASPT2 for $\zeta=50,5000$ and the limit $\zeta\to\infty$ are shown in
\Cref{fig:LiF_2states_z50,fig:LiF_2states_z5000,fig:LiF_2states_zInf}.
The off-diagonal elements of the Fock operator for $\zeta\to\infty$ in the three-state case (see main
text), are reported in \Cref{fig:LiF_zInf_f}. Note that the order of magnitude is the same as the
other XDW-CASPT2 calculations presented in the main text.
The PECs for $\zeta=500$ are presented in \Cref{fig:LiF_z500} and \Cref{fig:LiF_2states_z500} for
calculation with 3 and 2 states, respectively.
Density weights and off-diagonal Fock elements are shown in \Cref{fig:LiF_z500_w,fig:LiF_z500_f} for
the 3-state calculation with $\zeta=500$.

\clearpage
\section{Conical intersection of allene}

In this section we report additional computational details of the calculation involving the
distorted allene molecule that complement the description in the main text.

The total number of points in the two-dimensional scan is 6561, 81 for each variable.
The scan of the C-C-C-H torsion angle practically corresponds to a pyramidalization of the external
carbon atoms. Note that in the original work\cite{Granovsky2011}, the assignment of ``variable 1''
and ``variable 2'' in the surface plots were swapped as compared to the explanation in the text.

The IPEA shift was always set to zero for all methods.
For some cases it was necessary to include a real shift in order to converge the
calculation for all points.
In particular the XMS-CASPT2 calculation used a real energy shift of 0.5 Hartree for any size of the
model space.
The XDW-CASPT2 calculations were performed with a real energy shift of 0.5 Hartree in order to be
fairly compared to the XMS-CASPT2 results, even though a smaller shift was sufficient for converge
avoiding intruder states.

In order to have visually comparable results to Ref. \citenum{Granovsky2011}, we plotted
the absolute values of the energy difference between the $1^1$A$'$ and $2^1$A$'$ states
as a colormap with 64 contour iso-surface lines, using a color palette similar to the one of
the original work.

\begin{figure}
  \centering
  \includegraphics{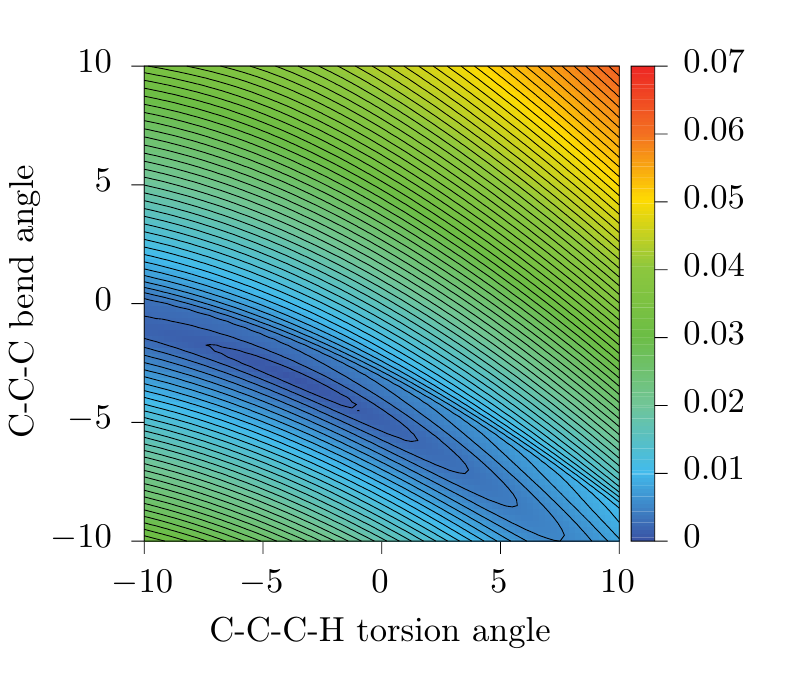}
  \caption{Colormapped isosurface plot of the absolute energy difference (in $\Eh$) between
  the $1^1$A$'$ and $2^1$A$'$ states for a model space including 6 states.
  The calculation was carried out with XMS-CASPT2.}
  \label{fig:allene_XMS-CASPT2_6x6}
\end{figure}

In \Cref{fig:allene_XMS-CASPT2_6x6} we report the results for a model space with 6 states for
XMS-CASPT2. As discussed in the main text, the PES remains virtually unchanged upon increase of
the model space dimension.

\clearpage
\section{Vertical excitation energies}

In this section we report additional information regarding the computational details for the
calculation of vertical excitation energies.
Imposed molecular point group symmetry, number of active electron, occupied and active orbitals
and real denominator shift are listed in \Cref{tab:comp_details_transitions}.
Note that if a shift was necessary for a method, it was also applied to the other ones in order
to obtain comparable results.
\begin{table}
    \captionsetup{width=\textwidth}
    \centering
    \caption{Computational details of CASSCF and CASPT2 calculations. The string of numbers
    in the occupied and active orbitals follows the irrep convention of OpenMolcas.}
    \begin{tabular}{lccllc}
    \toprule
    Molecule & PG & $N_{el}$ & Occ. MOs & Active MOs & Shift \\
    \midrule
    acetamide & C$_s$ & 6 & 12 1 & 1 3 & 0.1 \\
    acetone & C$_{2v}$ & 6 & 7 4 1 1 & 2 1 0 2 & 0.0 \\
    adenine & C$_s$ & 12 & 29 0 & 0 10 & 0.1 \\
    benzene & C$_s$ & 6 & 18 0 & 0 6 & 0.0 \\
    benzoquinone & D$_{2h}$ & 12 & 8 0 0 3 0 4 7 0 & 0 3 1 1 1 1 0 3 & 0.1 \\
    butadiene & C$_{2h}$ & 4 & 7 0 0 6 & 0 2 2 0 & 0.0 \\
    cyclopentadiene & C$_{2v}$ & 4 & 9 1 0 6 & 0 2 2 0 & 0.1 \\
    cytosine & C$_s$ & 10 & 24 0 & 0 8 & 0.0 \\
    formaldehyde & C$_{2v}$ & 4 & 5 0 0 1 & 0 2 0 1 & 0.0 \\
    formamide & C$_s$ & 6 & 9 0 & 1 3 & 0.1 \\
    furan & C$_{2v}$ & 6 & 9 0 0 6 & 0 3 2 0 & 0.0 \\
    hexatriene & C$_{2h}$ & 6 & 10 0 0 9 & 0 3 3 0 & 0.0 \\
    imidazole & C$_s$ & 8 & 14 0 & 1 5 & 0.1 \\
    naphthalene & D$_{2h}$ & 10 & 9 0 6 0 0 7 0 7 & 0 2 0 3 2 0 3 0 & 0.2 \\
    norbornadiene & C$_{2v}$ & 4 & 9 6 3 5 & 1 1 1 1 & 0.3 \\
    octatetraene & C$_{2h}$ & 8 & 13 0 0 12 & 0 4 4 0 & 0.0 \\
    propanamide & C$_s$ & 6 & 15 2 & 1 3 & 0.1 \\
    pyrazine & D$_{2h}$ & 10 & 5 0 0 3 0 4 4 0 & 1 2 1 0 1 0 1 2 & 0.2 \\
    pyridazine & C$_{2v}$ & 10 & 9 7 0 0 & 1 1 3 3 & 0.0 \\
    pyridine & C$_{2v}$ & 6 & 11 0 0 7 & 0 4 2 0 & 0.2 \\
    pyrimidine & C$_{2v}$ & 10 & 10 6 0 0 & 1 1 2 4 & 0.1 \\
    pyrrole & C$_{2v}$ & 6 & 9 0 0 6 & 0 3 2 0 & 0.0 \\
    tetrazine & D$_{2h}$ & 14 & 5 0 0 2 0 3 4 0 & 1 2 1 1 1 1 1 2 & 0.1 \\
    thymine & C$_s$ & 12 & 27 0 & 0 9 & 0.1 \\
    triazine & C$_s$ & 12 & 15 0 & 3 6 & 0.0 \\
    uracil & C$_s$ & 10 & 24 0 & 0 8 & 0.0 \\
    \bottomrule
    \end{tabular}
    \label{tab:comp_details_transitions}
\end{table}

\bibliography{references}